\documentclass[10pt]{article}

\usepackage{amsmath}
\usepackage{array}
\usepackage{appendix}
\usepackage{graphicx}
\usepackage{amsfonts}
\usepackage{amssymb}
\usepackage{mathrsfs}
\usepackage{yfonts}
\usepackage{euscript}
\usepackage{upgreek}
\usepackage{slantsc}
\usepackage{calligra}
\usepackage[T1]{fontenc}
\usepackage{epsf}
\usepackage{latexsym}

\usepackage{tipa}

\def\be{\begin{equation}}
\def\ee{\end{equation}}
\def\beq{\begin{equation}}
\def\eeq{\end{equation}}
\def\bea{\begin{eqnarray}}
\def\eea{\end{eqnarray}}

\def\ni{\noindent}
\def\foo{\footnote}

\def\hat{\widehat}

\def\!{\hspace{-1.6667em}}

\def\mD{\mbox{D}}
\def\mE{\mbox{E}}
\def\mF{\mbox{F}}

\def\mJ{\mbox{J}}  
\def\mK{\mbox{K}}

\def\mM{\mbox{M}}
\def\mN{\mbox{N}}

\def\mR{\mbox{R}}
\def\mS{\mbox{S}}

\def\me{\mbox{e}}

\def\mh{\mbox{h}}

\def\mp{\mbox{p}}

\def\ms{\mbox{s}}

\def\bupSigma{\mbox{\boldmath$\Sigma$}}                 

\def\fF{\mbox{\sffamily F}}

\def\fW{\mbox{\sffamily W}}

\def\ux{\underline{{x}}}

\def\bh{\underline{\underline{\mbox{h}}}  }            

\def\bM{\mbox{\bf M}}

\def\bM{\mbox{{\bf M}}}
\def\bM{\mbox{{\bf M}}}

\def\bh{\mbox{{\bf h}}}

\def\scA{\mbox{\scriptsize ${\cal A}$}}
\def\scB{\mbox{\scriptsize ${\cal B}$}}
\def\scC{\mbox{\scriptsize ${\cal C}$}}          
\def\scD{\mbox{\scriptsize ${\cal D}$}}          
\def\scE{\mbox{\scriptsize ${\cal E}$}}          
\def\scF{\mbox{\scriptsize ${\cal F}$}}
\def\scG{\mbox{\scriptsize ${\cal G}$}}          
\def\scH{\mbox{\scriptsize ${\cal H}$}}          
\def\scI{\mbox{\scriptsize ${\cal I}$}}
\def\scJ{\mbox{\scriptsize ${\cal J}$}}          

\def\scL{\mbox{\scriptsize ${\cal L}$}}          
\def\scM{\mbox{\scriptsize ${\cal M}$}}          
\def\scN{\mbox{\scriptsize ${\cal N}$}}
\def\scO{\mbox{\scriptsize ${\cal O}$}}
\def\scP{\mbox{\scriptsize ${\cal P}$}}
\def\scQ{\mbox{\scriptsize ${\cal Q}$}}          
\def\scR{\mbox{\scriptsize ${\cal R}$}}          
\def\scS{\mbox{\scriptsize ${\cal S}$}}

\def\scU{\mbox{\scriptsize ${\cal U}$}}          

\def\iB{\mbox{\scriptsize$B$}}   
\def\iD{\mbox{\scriptsize$D$}}   
\def\iK{\mbox{\scriptsize$K$}}   

\def\FrQ{\mbox{\Large $\mathfrak{q}$}}

\def\FrF{\mbox{$\mathfrak{F}$}}                                 
 
\def\FrM{\mbox{\Large $\mathfrak{m}$}}                         
\def\FrMgen{\mbox{\boldmath$\mathfrak{M}$}}                     

\def\sFG{\mbox{$\mathfrak{g}$}}

\def\FrG{\mbox{\Large $\mathfrak{g}$}}                            
 
\def\FS{\mbox{\LARGE\tt s}}                         

\def\sa{\mbox{\scriptsize a}}

\def\scc{\mbox{\scriptsize c}}
\def\sd{\mbox{\scriptsize d}}
\def\se{\mbox{\scriptsize e}}

\def\sg{\mbox{\scriptsize g}} 
 
\def\si{\mbox{\scriptsize i}}

\def\sll{\mbox{\scriptsize l}}  
\def\sm{\mbox{\scriptsize m}}
\def\sn{\mbox{\scriptsize n}} 
\def\so{\mbox{\scriptsize o}}

\def\sr{\mbox{\scriptsize r}}
\def\st{\mbox{\scriptsize t}}

\def\sw{\mbox{\scriptsize w}}

\def\sA{\mbox{\scriptsize A}} 
\def\sB{\mbox{\scriptsize B}}

\def\sF{\mbox{\scriptsize F}}
\def\sG{\mbox{\scriptsize G}}

\def\sJ{\mbox{\scriptsize J}}
\def\sK{\mbox{\scriptsize K}}
 
\def\sM{\mbox{\scriptsize M}} 
\def\sN{\mbox{\scriptsize N}} 

\def\sP{\mbox{\scriptsize P}} 
 
\def\sR{\mbox{\scriptsize R}}
\def\sS{\mbox{\scriptsize S}}

\def\sW{\mbox{\scriptsize W}}

\def\sfA{\mbox{\sffamily{\scriptsize A}}}      
\def\sfB{\mbox{\sffamily{\scriptsize B}}}      
\def\sfC{\mbox{\sffamily{\scriptsize C}}}      
\def\sfD{\mbox{\sffamily{\scriptsize D}}}      
\def\sfF{\mbox{\sffamily{\scriptsize F}}}      
\def\sfG{\mbox{\sffamily{\scriptsize G}}}      
\def\sfI{\mbox{\sffamily{\scriptsize I}}}      
\def\sfK{\mbox{\sffamily{\scriptsize K}}}      
\def\sfQ{\mbox{\sffamily{\scriptsize Q}}}      

\def\sbM{\mbox{{\bf \scriptsize M}}}

\def\tfA{\mbox{\sffamily{\tiny A}}}

\def\tfW{\mbox{\sffamily{\tiny W}}}

\def\K{Kucha\v{r} }

\def\CPI{Conditional Probabilities Interpretation }

\def\pa{\partial}
\def\d{\textrm{d}}

\def\5Star{\mbox{\Large$\star$}}              
\def\cr{\mbox{\scriptsize{\bf $\mbox{ } \times \mbox{ }$}}}

\def\sumi2{\sum\mbox{}_{\mbox{}_{\mbox{\scriptsize $i$=1}}}^2}
\def\sumi3{\sum\mbox{}_{\mbox{}_{\mbox{\scriptsize $i$=1}}}^3}

\def\sumj3{\sum\mbox{}_{\mbox{}_{\mbox{\scriptsize $j$=1}}}^3}
\def\sumk3{\sum\mbox{}_{\mbox{}_{\mbox{\scriptsize $k$=1}}}^3}

\begin{document}

\begin{titlepage}

\begin{center}

\Huge{\bf WHERE TO APPLY RELATIONALISM}


{\large \bf Edward Anderson} 

\vspace{.15in}

\large {\em DAMTP, Centre for Mathematical Sciences, Wilberforce Road, Cambridge CB3 OWA.} \normalsize

\end{center}

\begin{abstract}

Relationalism -- along the lines developed by Barbour and collaborators in the past 3 decades -- 
can be considered an advance with 1/4 of the facets of the canonical approach's Problem of Time as identified by Isham and Kucha\v{r}. 
Indeed, almost all of the Problem of Time facets have classical counterparts, 
since they arise from consequences of demanding background independence rather than about combining GR and QM per se. 
Moreover the quantum version is harder, while the classical counterpart provides some suggestions through being more solvable. 
The suggestion then is to consider the effect of this advance on the Problem of Time as a whole, 
as opposed to repeating the same classical portion for different redundancy groups acting on the configuration space: shape dynamics. 
There are indeed some knock-on effects because the facets are notoriously not independent.   
The other facets do also however require distinct insights. 
Finally, I comment on the above being `metric' background independence, whereas quantum gravity usually assumes many other levels of background structures. 
Whilst Isham already wrote about this over two decades ago, 
it has largely not yet been incorporated into quantum gravity programs, and could well be a good area to extend and re-envigour by use of relational thinking. 

\end{abstract}

\section{Introduction} 

\ni {\it Sagredo}: Let us talk about Quantum Gravity. 
This is the end point of combining three theories, each of which is characterized by a fundamental constant: 
$\hbar$ from Quantum Mechanics, $1/c$ from Special Relativity and $G$ from Gravitation.  
One then considers theories involving pairs of these.
Finally, Quantum Gravity itself is characterized by the Planck units that combine all three of these, thus completing the cube of theories in Fig 1.a).   

\ni {\it Salviati}: In some senses this is so.
However, upon taking the edges of your cube to be maps between theories, one can readily envisage that these maps may not commute: Fig 1.b). 
Thus your use of `the' in `the endpoint' is questionable.  
Furthermore, the nature of the quantities that the units correspond to also raises issues.
In particular, the theories represented on your cube lie within distinct paradigms which attribute incompatible meanings to the word `time'...

\ni {\it Sagredo}:  And by this you mean the incompatibility between Quantum Mechanics and General Relativity (GR) that constitutes the notorious Problem of Time?  

\ni {\it Salviati}: Not quite. 
The Problem of Time \cite{Kuchar92, I93} is often described in terms of such an incompatibility.
However, the Problem of Time is more generally a consequence of \cite{BI, APoT3} demanding Background Independence \cite{A6467Giu06}. 
The incompatibility in question can thus be traced back to being between background dependent theories and background independent ones.
For instance, such an incompatibility already occurs at the classical level...
Indeed, one can consider classical GR as not just a Relativistic Theory of Gravitation but furthermore as a gestalt of that and of a freeing from background structure.
That is in line with Mach and Einstein's perspectives on Physics. 
One can even consider Mechanics in this manner, in that Relational Particle Mechanics \cite{BB82, FileR} has a number of background-independent features.

\ni {\it Sagredo}: But is the Problem of Time not the frozenness of the Wheeler--DeWitt equation of canonical quantum GR? 

\ni {\it Salviati}: Firstly, Relational Particle Mechanics' quantum equations also look frozen, so the feature you mention is not exclusive to GR or even to Gravitation. 
Secondly, the canonical approach's Problem of Time has been argued to have eight facets \cite{Kuchar92, I93}. 
Additionally, other approaches based on spacetime and path integrals involve a ninth facet instead of some parts of the aforementioned eight. 
Thus the Frozen Formalism Problem is in fact only a small part of the Problem of Time.

\ni {\it Sagredo}: But we can look at this part of the picture first?

\ni {\it Salviati}: To some extent one can do so, at the conceptual level. 
However the Problem of Time's facets interact nonlinearly toagreat extent, as was already pointed out long ago \cite{Kuchar92, I93, Kuchar93} 
(see also \cite{APoT, APoT2, ABook} for updates).

\ni {\it Sagredo}: But Physics projects work by solving one problem at a time, building up on existing solutions? 

\ni {\it Salviati}: Sometimes, but very largely not for this particularly thorny foundational problem.

\ni {\it Sagredo}: One point I intended to make in mentioning the Frozen Formalism Problem is that it appears to be phrased as a quantum-level problem...

\ni {\it Salviati}: In fact, Sec 2 of this Seminar argues that this problem follows naturally from a type of Background Independence called Temporal Relationalism.  
This leads to GR's Hamiltonian constraint $\scH$ whose form then gives the Frozen Formalism Problem at the quantum level.
Thus the often-used phrasing for the Frozen Formalism Problem is not evidence against the Problem of Time already being an issue at the classical level... 

\ni {\it Sagredo}: The other point I wished to make is to ask how specific is the Problem of Time to GR,  
and indeed to the rather antique geometrodynamical formulation of this that it is so often presented for?

\ni {\it Salviati}: This is a good point to raise! 
The geometrodynamical formulation indeed dates back to the 1960's 
(well-known works of Arnowitt, Deser and Misner \cite{ADM}, Dirac \cite{Dirac}, Wheeler \cite{Battelle} and DeWitt \cite{DeWitt67}).
And by the 1980's most work in 

\end{titlepage} 

{            \begin{figure}[ht]
\centering
\includegraphics[width=0.8\textwidth]{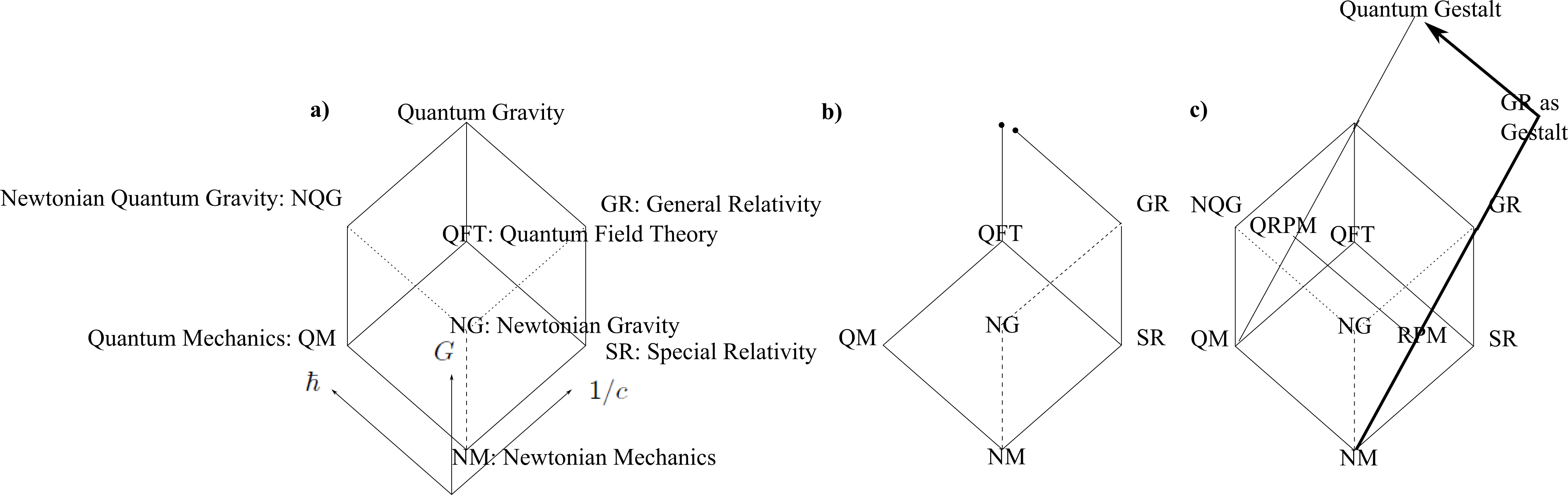}
\caption[Text der im Bilderverzeichnis auftaucht]{        \footnotesize{a) Planckian cube of fundamental physical theories.  
b) indicates the Newton--Einstein (= classical = non-quantum) plane and the Particle Physics (= non-gravitational) plane.   
c) Gordian cube: here it is perceived that different routes along the edges to the `final Quantum Gravity vertex' may not commute (in the algebraic sense, upon viewing its edges as maps).
d) Cutting the Gordian cube, or `thinking outside of the box' that is a).}  }
\label{Cubes-3}\end{figure}          }

\ni Quantum Gravity had moved on to other approaches.
Furthermore, the geometrodynamical formulation is indeed often associated with the Problem of Time since it certainly manifests this 
and is indeed the specific example used in many of the Problem of Time's traditional and well-known papers \cite{Kuchar92, I93, B94I}. 
However, the Problem of Time is not just a problem encountered in the geometrodynamical approach, 
but is rather a pervasive problem that so happens to already be well-illustrated by its well-known geometrodynamical subcase.
So for instance, these issues continue to apply in the more modern Loop Quantum Gravity approach, which indeed also places importance on Background Independence... 
Also, M-Theory is background independent and Supergravity can be considered in that light too. 
Consequently the Problem of Time manifests itself in each of these cases as well.

\ni {\it Sagredo}: Thank you for that clarification. 
Returning to your ante-penultimate answer, does this mean that the subject of this Conference -- Barbour's shape dynamics \cite{Kos2, Mercati14} -- 
is the way forward as regards tackling the Problem of Time?  

\ni {\it Salviati}: The Temporal Relationalism I mention above, alongside a second aspect of Background Independence -- Configurational Relationalism -- 
indeed originated from works of Barbour and collaborators in the 1982--2002 period \cite{BB82, RWR}.
[Configurational Relationalism involves a physically irrelevant group $\FrG$ acting upon a system's configurations. 
This leads to constraints that are linear in the momenta.  
In the case of GR, $\FrG$ is the 3-diffeomorphisms, and the corresponding constraint is the momentum constraint $\scM_i$.]
In particular, the maturation of this Relational Approach from 1994 \cite{B94I} until 2002 
provides useful conceptual steps forward from the position given in the well-known Problem of Time reviews \cite{Kuchar92, I93}.
Indeed, Temporal and Configurational Relationalism constitute very general and already classically relevant Background Independence underpinnings 
for two of the nine Problem of Time facets \cite{APoT3}.  
Shape dynamics then involves seeking to redo this work for different $\FrG$. 
This is certainly interesting, but what about the distinct program of extending Background Independence underpinning to covering the other seven Problem of Time facets?
These are the two chief relational systems in the present literature that both follow from Barbour's work in the 1982 until 2002 period.
And this Seminar lays out the argument for the second of these programs.

\ni {\it Sagredo}: That is certainly also interesting!
What happens in this approach?   

\ni {\it Salviati}: It demonstrates that the Problem of Time Facets are indeed rooted in Background Independence assumptions, which are conceptually and philosophically desirable.
It establishes how Barbour's brand of Relationalism (Temporal and Configurational) can be enlarged to a more general theory of Background Independence \cite{BI, APoT3}.  
Indeed GR can be recast in relational form.
Historically and logically this is via Arnowitt, Deser and Misner's \cite{ADM}, and Baierlein, Sharp and Wheeler's \cite{BSW} works, 
Wheeler's interpretational considerations \cite{WheelerGRT, Battelle}, 
then Barbour's consolidation \cite{B94I} of his earlier approach to Physics with Bertotti \cite{BB82}, and various subsquent workings and improved conceptualizations \cite{RWR, AM13}.
By this stage, GR can be re-dervived from relational premises in which 3-space is primary, 
and one considers which of the many imaginable theories of geometrodynamics are consistent... 

\ni {\it Sagredo}: But what of GR's conventional spacetime formulation? 

\ni {\it Salviati}:  Wheeler \cite{Battelle} provided convincing arguments against the fundamentality of spacetime at the quantum level; see Sec \ref{SCP} for an outline.
Quantum Mechanics in fact unfolds on configuration space (or some other polarization half-set of a physical theories position and momentum variables.
And indeed, on the other hand, the Barbour-type Relational approaches take configuration spaces to be primary entities.
Moreover, upon re-deriving GR from relational premises, it is {\sl found} to possess spacetime structure at the classical level. 
Indeed, various further desirable aspects of Background Independence {\sl follow from} the emergence of this spacetime structure... 

\ni {\it Sagredo} But surely then these aspects were already known from the traditional spacetime-first approach to GR?

\ni {\it Salviati} That is so, 
once one's knowledge includes the Arnowitt--Deser--Misner split and its subsequent conceptualization in terms of embeddings, slices and foliations \cite{T73}.
Moreover, the Background Independence analysis of which I speak leads to recognizing more than the traditional number of Problem of Time facets! 
For instance, \cite{APoT3} provides in this manner a new list of {\sl twenty-five} facets...

\ni {\it Sagredo}: So you can {\sl more detailedly characterize} the Problem of Time, as well as following its roots back to Background Independence.
But does this {\sl improve the prospects of solving it}?

\ni {\it Salviati}: The Background Independence and Problem of Time analysis can indeed be done at the classical and semiclassical levels 
for a range of model arenas going as far as realistic perturbatively inhomogeneous Quantum Cosmology.  
Moreover, each of the classical, semiclassical and fully quantum levels is progressively harder to handle.

\ni {\it Sagredo}: Let us now return the shape dynamics horn of the argument, since it is the main theme of this Conference.
Can you say more about how this arises in relation to the Background Independence program?

\ni {\it Salviati} Let me first clarify that a third Problem of Time facet concerns whether the set of constraints that one is aware of suffice, 
or whether they give rise to further constraints.
This is determined by a type of Dirac procedure \cite{Dirac}; this is what I meant before when I said `consistent'.
Now in the relational approach, Temporal and Configurational Relationalism {\sl provide the constraints} $\scH$ and $\scM_i$ respectively, 
so the issue then is whether these close algebraically. 
Furthermore, if one is not presupposing GR, $\scH$ is replaced by a family of candidate entities, and the question is which of these are consistent. 
Wheeler had already asked whether there were first principles leading to the specific form of GR's $\scH$. 
The relational first principles are then an interesting possibility for this, with the Dirac approach acting as a filter.  
What then happens is that an obstruction term with four factors arises from the algebraic structure formed by the constraints \cite{RWR, AM13}. 
This can be overcome by whichever of its factors vanishing. 
In this manner, one of these factors gives back locally Lorentzian GR with its conventional spacetime structure. 
Two of the other factors give the infinite maximum signalling speed limit of GR with its Galilean relativity and the opposite zero maximum signalling speed limit 
as envisaged in `Alice through the looking glass'. 
It is very satisfying that these choices of local Theory of Relativity arise algebraically from Dirac's procedure for handling constraints! 
Moreover, the obstruction term associates these three choices with an unexpected fourth choice via the form of the last factor.
This factor vanishes if the spatial slice is of constant mean curvature, which is well known to be associated with spatially conformal mathematics \cite{YorkTime1}. 
This is how the shape dynamics option first arose.
Moreover, its equations can then be built in from the start \cite{ABFKO, Kos2} by redoing Configurational Relationalism with a different group $\FrG$.  
Yet as I shall explain in this Seminar, the other chief relational system adheres to the first and not fourth prong of this fork, 
since the first incorporates {\sl yet further} aspects of Background Independence... 

\ni {\it Sagredo}: You describe shape dynamics as redoing with different $\FrG$'s.  
But is shape dynamics {\sl just} redoing the same procedure?
Are the conformal transformations adjoined to its its $\FrG$ not rather special and significant? 

\ni {\it Salviati}: There is a sense in which they are, and another sense in which they are but one step where many more could be carried out.
This is most straightforwardly explained by laying out the layers of mathematical structure used in Physics (Fig \ref{Bigger-Set-2}).
Then indeed considering conformal transformations to be physically irrelevant corresponds to a geometrically well-defined level of mathematical structure 
in between Riemannian geometry and differential geometry. 
On the other hand, why cease to consider Relationalism, Background Independence and the Problem of Time at this particular level? 
The Background Independence program is open to looking all the way down: which aspects transcend all the levels and which cease to apply 
(or be defined or distinct) below a particular level.
Thus the Relationalism, Background Independence and Problem of Time at the levels of topological manifolds, topological spaces and sets are further ports of call...  
Nor have I yet mentioned all known possibilities.
Firstly, let us return to the cube of theories. 
This is in the sense that the theories in the cube have common first variants that are conformal, 
thus further motivating conformality, but an alternative common first variant also exists: supersymmetic theories.
[In fact, the subsequent `topological' variant of the cube of theories is also quite widely known, though this refers to topological manifolds rather than topological spaces.]
So then what about the Relational and Background Independence statuses of supersymmetric theories and what subsequent Problem of Time facets these exhibit? 

\ni {\it Sagredo}: As regards your pointing to Supersymmetry, I agree that the conformal case does not stand alone in being interesting and a possible next case to study! 
On the other hand, as regards your mention of topological manifolds and beyond, is Background Independence not just taken to mean at the level of Metric Geometry?

\ni {\it Salviati}: Firstly, by involving diffeomorphisms, 
what is usually termed metric-level Background Independence is really at the level of metric {\sl and} differentiable structure... 
But yes, the current program in any case asks for more that various other existing programs 
in its consideration of the extent to which Relationalism and Background Independence continue to apply at the level of topological manifolds and below.
Moreover, the other programs stop at the metric level for {\sl convenience}, 
rather than out of {\sl having principles as regards why Background Independence should apply no further}. 
On the other hand, the Background Independence program involves \cite{ATop, AMech} delving level by level 
until if and when the physics and mathematics itself provides convincing first principles for why Background Independence applies no further.

\ni {\it Sagredo}: You make a good case to also study Relationalism more widely!
Your program complements the other well and provides future frontiers for it...   
I hope that the shape dynamics people work on this too, lest they risk over-specializing!
But how far down the chain has such thinking been applied so far? 
 
\ni {\it Salviati} At the quantum level, the Academician not only pioneered these further steps long ago \cite{I89-Latt, IKR-I91} 
but also more recently considered these for alternative foundational systems of mathematics \cite{I03, I10-ToposRev}.  
The latter involve a second sense in which Fig \ref{Bigger-Set-2} is not all-seeing, 
namely that it is based on the most common mathematical foundational system of equipping sets with various further structures.
However, this is not the only way of doing mathematics; others are along the lines of Category Theory, Topos Theory and Synthetic Mathematics. 
The present Seminar, however, is confined to Fig \ref{Bigger-Set-2}'s paradigm of mathematics alongside a brief account of the supersymmetric option.
Simpler considerations \cite{ATop} reveal the classical domain to already possess nontrivial counterparts for the range of steps down the conventional levels of 
mathematical structure that were pioneered at the quantum level in \cite{I89-Latt, IKR-I91}.  
Sec \ref{Deeper} provides a brief outline of this! 
%

\section{The nine aspects leading to the nine facets}

As a first dichotomy, consider dynamical primality versus spacetime primality. 
In the dynamical primality alternative, one can start with configuration space (a generalization of the spatial primality position).  
Then a second dichotomy is between Solipsism and approaches allowing change.

{            \begin{figure}[ht]
\centering
\includegraphics[width=0.40\textwidth]{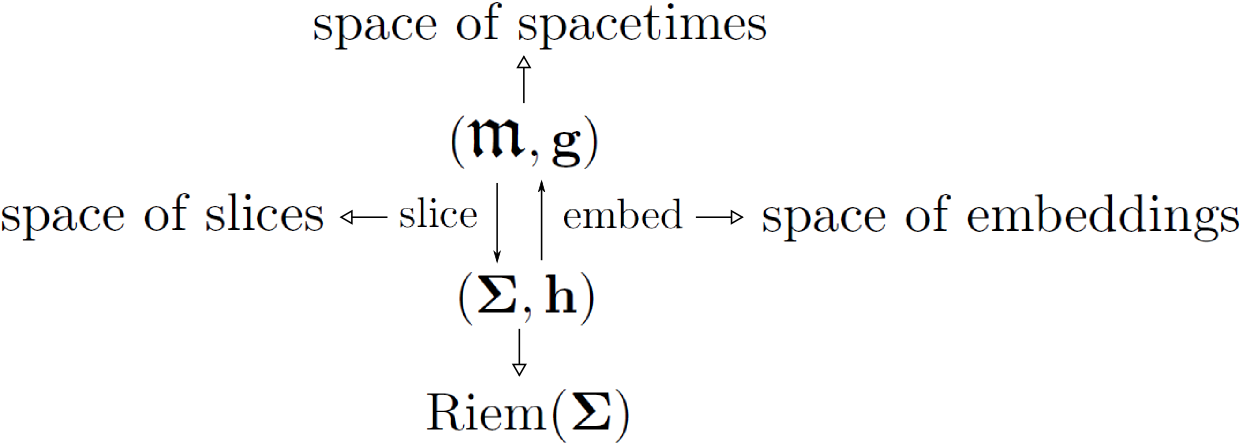}
\caption[Text der im Bilderverzeichnis auftaucht]{\footnotesize{One passes from spacetime to space by considering a slice and projecting spacetime entities onto it, 
or by foliating the spacetime with a collection of spaces.
Moving in the opposite direction involves embedding rather than projecting, and is a construction (which is more involved due to assumption of less structure).  
Compare this two-way passage with two of Wheeler's routes and with Problem of Time (PoT) Facets 6) and 7) below.   
Moreover, one is now to interpret spacetime versus space as a dichotomy of primality; the case for space primality is a subcase of that for the primality of dynamics.
Including the spaces of each of the four preceding entities (linked by white arrows), 
one arrives at the eightfold that is crucial for understanding many of the facets of the PoT.} }
\label{Crux-Plain}\end{figure}            }

\subsection{Temporal Relationalism and the Frozen Formalism Problem}

%
\ni {\it Temporal Relationalism} concerns there being no time for the universe as a whole. 
This is a theory-independent statement, holds just as well at the classical level than at QM level.  
It is useful via admitting a mathematically sharp implementation, e.g. as follows.

\mbox{ } 

\ni i) This action is not to include any extraneous times (such as $t^{\sN\se\sw\st\so\sn}$) or extraneous time-like variables (such as the ADM lapse of GR, $\upalpha$).

\ni ii) Time is not to be smuggled into the action in the guise of a label either.

\mbox{ }  

%
\ni One formulation of ii) is for a label to be present but physically meaningless because it can be changed for any other (monotonically related) label 
without changing the physical content of the theory.   
I.e. the action in question is to be {\it manifestly reparametrization invariant}. 

\ni The classical-level Jacobi formulation follows from the action principle   

\ni \beq
\FS_{\sJ} := \int \d\lambda \, L_{\sJ} := 2 \int \d\lambda \sqrt{T W} \mbox{ } .  
\label{S-Jacobi}
\eeq
For later reference, it is useful to envisage ${\d}/{\d\lambda}$ as the Lie derivative $\pounds_{{\d}/{\d\lambda}}$ in a particular frame \cite{Stewart}. 
In the Mechanics case, the kinetic energy $T := ||\dot{\mbox{\boldmath{$q$}}}||_{\mbox{\scriptsize\boldmath{$m$}}}\mbox{}^2/2$ where the configuration space metric $\mbox{\boldmath{$m$}}$ is just the 
`mass matrix' with components $m_{I}\delta_{IJ}\delta_{ij}$, and with {\it potential factor} $W := E - V(\mbox{\boldmath{$q$}})$ for $V(\mbox{\boldmath{$q$}})$ the potential energy and 
$E$ is the total energy of the model universe.  
In the particular minisuperspace case that I consider in this article, $TW$ is of the form 

\ni\beq
\mbox{exp}(6\Omega)\left\{- \left\{\frac{\d\Omega}{\d \lambda}\right\}^2 + \left\{\frac{\d\phi}{\d \lambda}\right\}^2\right\}\big\{\mbox{exp}(-2\Omega) - V(\phi) - 2\Lambda\big\}                                                                       \mbox{ } . 
\label{MSS-Action}
\eeq
Here, the {\it Misner variable} $\Omega := \mbox{ln} \, a$, for $a$ the usual cosmological scale factor. 
N.B. that in each case the meaningless $\lambda$'s cancel out in the action.

Reparametrization Invariance necessarily \cite{Dirac} provides a primary constraint; in the GR case, this is the Hamiltonian constraint $\scH$. 
On the other hand, in the Mechanics case, it is an `energy' constraint, $\scE$ 

\ni \beq
\scE := ||\mbox{\boldmath{$p$}}||_{\mbox{\scriptsize\boldmath{$n$}}}/2 + V(\mbox{\boldmath{$q$}}) = E \mbox{ } .  
\label{E-Constraint}
\eeq
Here $\mbox{\boldmath{$n$}} = \mbox{\boldmath{$m$}}^{-1}$, with components $\delta_{IJ}\delta_{ij}/m_I$.  
Next note that in these cases, the {\it purely quadratic form}  of the above relational action (in the velocities) is inherited by the primary constraint (in the momenta).
The GR case of this is well-known to lead to the quantum-level {\it Wheeler--DeWitt equation} \cite{DeWitt67, Battelle}

\ni\be
\widehat\scH\Psi = 0 \mbox{ } 
\ee
for $\Psi$ is the wavefunction of the universe.
This manifests frozenness via being a subcase of time-independent Schr\"{o}dinger equation (TISE) $\widehat{H}\Psi = E\Psi$. 
I.e. a stationary alias timeless alias frozen wave equation.
Furthermore, this occurs in a situation in which one might expect a time-dependent Schr\"{o}dinger equation $\widehat{H}\Psi = i\hbar \, \pa\Psi/\pa t$ for some notion of time $t$.

In more detail, the Mechanics case is along the lines of $\widehat{\scE}\Psi = -\frac{\hbar^2}{2}\triangle\Psi + V\Psi = E\Psi$.
The GR case is\foo{Here `\mbox{ }' implies in general various well-definedness, regularization and operator-ordering issues.} 

\ni\beq
\widehat{\scH}\Psi := - \hbar^2`\frac{1}{\sqrt{\mM}}  \frac{\delta}{\delta \mh^{\mu\nu}}
\left\{
\sqrt{\mM}\mN^{\mu\nu\rho\sigma}  \frac{\delta}{\delta \mh^{\rho\sigma}}
\right\} 
+ \sqrt{\mh}\{2\Lambda - \mR(\ux; \bh]\}'\Psi + \hat\scH^{\mbox{\scriptsize matter}}\Psi = 0   \mbox{ } . 
\label{WDE} 
\eeq
Note that this (and for field theories more generally) contains in place of a partial derivative $\pa/\pa Q^{\sfA}$ a functional derivative $\delta/\delta h_{ij}(x^k)$.  

\ni Comment 1) Is there a paradox between the Temporal Relationalism Postulate and our appearing to `experience time'?  
To begin to form an answer, `time' is a useful concept for everyday experience. 
However the nature of this time concept is less clear, and also everyday experience is of subsystems rather than concerning the whole universe.

\ni Comment 2) Moreover, we have made clear that this is not a quantum-level surprise, but rather inherited from classical Temporal Relationalism.
We next consider how to resolve Temporal Relationalism's ab initio timelessness for the universe as a whole at the classical level.
This marks the start of this Seminar's ordered passage through the various PoT facets, which can be picked out in the text by its enumeration by `steps'.  

\ni Step 1) To counter Temporal Relationalism, recollect Mach's Time Principle \cite{M}: that `time is to be abstracted from change'. 
I.e. timelessness for the universe as a whole at the primary level is {\sl resolved} by time emerging from change at the secondary level from Mach's Time Principle. 
To best access this, upgrade the implementation of Temporal Relationalism as per the following reformulations \cite{BI, AM13, ABook}.  

\ni It is a further conceptual advance to formulate one's action and subsequent equations without use of any meaningless label at all.  
I.e. a {\it manifestly parametrization irrelevant} formulation in terms of {\sl change}           $\d Q^{\sfA}$             rather than 
     a      manifestly reparametrization invariant          one in terms of a label-time velocity $\d Q^{\sfA}/\d \lambda$.  
Note that this casts the implementation of Temporal Relationalism explicitly in terms of {\sl change}: the resolving entity in the statement of Mach's Time Principle.

\ni Finally, it is better still to formulate this without even mentioning any meaningless label/parameter, 
by use of how the preceding implementation happens to be dual to a configuration space geometry formulation.\footnote{As to the truth and naturality of this, 
Jacobi's action principle is indeed often conceived of in these geometrical terms rather than in its dual aspect as a timeless formulation.}

\ni Step 2) The next issue is what type of change the emergent time is to involve.
Some changes are more significant than others, and there are some senses in which including more changes increases accuracy.
However, some changes are only poorly known in practise, so their inclusion would in other senses decrease accuracy. 
Because of this, I do not posit using all change, but rather a {\it sufficient totality of locally relevant change (STLRC)} \cite{ARel2, ABook}. 
Both the `all change' and STLRC positions are conceptually along the lines of the astronomers' ephemeris time \cite{Clemence}.
However, it is the latter that most accurately reflects the astronomers' actual procedure in determining ephemeris time 
(which concentrates on solar-system bodies -- locally relevant -- rather than on all changes in the universe).

\ni Step 3) A specific classical-level implementation of a Machian emergent time is then as follows. 
It is distinguished by its simplification of the model's momentum--velocity relations and equations of motion using 
$\pa/\pa t^{\se\sm(\sJ)} := \sqrt{W/T}\pa/\pa\lambda = \sqrt{2W}/\d s$.  
This can be integrated up to give

\ni\beq
t^{\se\sm(\sJ)} = \int \d s/\sqrt{2W}  \mbox{ } 
\label{t-em-J}
\eeq
(the J stands for Jacobi).
In principle, this includes `all change', but its operational specification allows for its actual computation to retain only the STLRC for the situation in question.  
Note that this equation is a rearrangement of $\scE$ (or, for now, the minisuperspace case of $\scH$).
In particular, this means that $\scE$ is here being interpreted as an {\it equation of time} rather than as an `energy constraint'.
Due to this, I denote the general such constraint by $\scC\scH\scR\scO\scN\scO\scS$.
Thus Temporal Relationalism leads to a primary constraint $\scC\scH\scR\scO\scN\scO\scS$, 
the significance of which is that of an equation of time, that can rearranged to give a timestandard emergent from change: an implementation of Mach's Time Principle. 

\ni Comment 3) This is very satisfying from the perspective of {\it constraint providers}: underlying physical and philosophical reasons for constraints.
More generally, having noted that e.g. GR is a constrained dynamical formulation, one can ask for underlying reasons why the theory of nature might have such constraints.
E.g. Wheeler \cite{Battelle} asked the following question, which readily translates to asking for first-principles reasons for the form of the 
crucially important GR Hamiltonian constraint, $\scH$:

\ni\beq
\stackrel{\mbox{\it \normalsize ``If one did not know the Einstein--Hamilton--Jacobi equation, how might one hope to derive it straight off from}} 
         {\mbox{\it \normalsize plausible first principles without ever going through the formulation of the Einstein field equations themselves?"}} 
\label{Wheeler-Q}
\eeq
The above is then part of an answer to this particular question of constraint provision (see Sec \ref{SCP} for the rest of this answer).   
The next Sec concerns a different type of constraint provider that leads e.g. to GR's momentum constraint $\scM_i$.
Contrast with a different attitude to constraints (from Applied Mathematics) is that one is simply prescribed a set of such ab initio. 
Finally, the reverse process to `provide' is `encode'.  
I.e. to subsequently build into one's action auxiliary variables variation with respect to which encodes the constraints; further reasons for encoding become apparent in Sec \ref{CC}.

\ni Comment 4) For relational particle mechanics (RPM), $t^{\se\sm}$ amounts to a relational recovery of Newtonian time.  
On the other hand, for the geometrodynamical formulation of GR, $t^{\se\sm}$ amounts to a recovery of GR proper time.

\ni Comment 5) Moreover, the above suite of implementations of Temporal Relationalism can be extended to cover {\sl every} level of structure rather than just at the level of Lagrangian 
(or Jacobian) formulations \cite{FEPI, TRiPoD, AM13}.
This is necessary in order for Temporal Relationalism to be tractable in tandem with other PoT facets \cite{TRiFol, ABook}.
{\sl Reformulating the Principles of Dynamics to be Temporal Relationalism implementing -- TRiPoD -- ensures that all subsequent considerations of other Problem of Time facets 
within this new paradigm do not violate the Temporal Relationalism initially imposed}. 
Thus one requires a reformulation of the entirety of the Principles of Dynamics to ensure one does not find oneself outside the `Frozen Formalism' gate again at some later stage in one's 
journey that attempts to resolve the PoT. 
One might do well at this point to take how one requires all that so as to remain safely inside {\sl just one} of the gates as an indication of the severity of the Problem of Time. 
In particular, a lot of work needs to be done in order to not lose one's program's previous successes with a subset of the gates when one tries to extend that program to pass through 
more of the gates.

\subsection{Configurational Relationalism, Best Matching and the Thin Sandwich Problem}

%
\ni This term covers both a) {\it Spatial Relationalism} \cite{BB82}: no absolute space properties.
b) {\it Internal Relationalism} is the post-Machian addition of not ascribing any absolute properties to any additional internal space that is associated with the matter fields.
Configurational Relationalism is addressed as follows.

\ni i) One is to include no extraneous configurational structures either (spatial or internal-spatial metric geometry variables that are fixed-background rather than dynamical).

\ni[Since time-parametrization is really a 1-$d$ metric of time, the i) condition of both Temporal and Configurational Relationalism reflect a single underlying relational 
conception of Physics: no fixed-background metric-level geometry.] 
 
\ni ii) Physics in general involves not only a $\FrQ$ but also a $\FrG$ of transformations acting upon $\FrQ$ that are taken to be physically redundant.
[This is a {\it group action}: in general a map $\FrG \times X \rightarrow X$ for group $\FrG$ acting on some other mathematical space $X$.]

\ni ii) is a matter of practical convenience: often $\FrQ$ with redundancies is simpler to envisage and calculate with.
Also for ii) the Internal Relationalism case is another formulation of Gauge Theory from that presented in books on QFT. 
The spatial case is similar, in that it can also be thought of as a type of Gauge Theory for space itself.
This includes modelling translations and rotations relative to absolute space as redundant in Mechanics, for which $\FrQ = \mathbb{R}^{Nd}$,   
                                                         or Diff($\bupSigma$) as redundant in GR,        for which $\FrQ = \mbox{Riem}(\bupSigma)$: the space of Riemannian 3-metrics.
See \cite{ABook} for restrictions on $\FrQ$, $\FrG$ pairings.  
				
\ni Implementing Configurational Relationalism at the level of Lagrangian variables ($\mbox{\boldmath $Q$}$, $\dot{\mbox{\boldmath $Q$}}$) is known as {\it Best Matching} \cite{BB82}.
This involves pairing $\FrQ$ with a $\FrG$ such that $\FrQ$ is a space of redundantly-modelled configurations.
Here $\FrG$ acts on $\FrQ$ as a shuffling group: one considers pairs of configurations, keeping one fixed and shuffling the other 
(i.e. an active transformation) until the two are brought into maximum congruence.

In more detail, one proceeds via constructing a $\FrG$-corrected Principles of Dynamics action.
For the examples considered here, this involves replacing each occurrence of $\dot{\mbox{\boldmath $Q$}}$ with 
$\dot{\mbox{\boldmath $Q$}} - \stackrel{\rightarrow}{\FrG_g}\mbox{\boldmath $Q$}$, where $\stackrel{\rightarrow}{\FrG_g}$ indicates group action.
Note that whereas these correction terms can be interpreted as fibre bundle connections (see e.g. \cite{Mercati14}), 
they can also be interpreted as Lie derivatives (see e.g. \cite{Stewart, FileR}).
The latter is more minimalistic since it is at the level of differential geometry without having to assume connection or bundle structures.

Then varying with respect to the $\FrG$ auxiliary variables $g^{\sfG}$ produces the {\it shuffle constraints} $\scS\scH\scU\scF\scF\scL\scE_{\sfG}$. 
These arise as {\sl secondary constraints}: via use of equations of motion, and are linear in the momenta.
In setting up Best Matching, the intent is that $\FrG$ acts as a gauge group,
though one only knows that intent has succeeded once one has obtained the brackets between the constraints [Aspect 3) below].
Thus $\scS\scH\scU\scF\scF\scL\scE_{\sfG}$ for now remains a {\sl candidate} constraint associated with an {\sl attempt} to associate $\FrG$ to $\FrQ$.  
As a separate second point, note that the initial introduction of $\FrG$ corrections appears to be a step in the wrong direction as regards freeing the physics of $\FrQ$ from $\FrG$.  
This is due to its extending the already redundant space $\FrQ$ of the $\mbox{\boldmath $Q$}$ to some joint space of $\mbox{\boldmath $Q$}$ and explicitly-mentioned $g^{\sfG}$.  
However, if the candidate is successful, the end-product of the next step $\scS\scH\scU\scF\scF\scL\scE_{\sfG}$ is of the form $\scG\scA\scU\scG\scE_{\sfG}$, 
which (Sec \ref{CC}) is a type of constraint that uses up {\sl two} degrees of freedom per $\FrG$ degree of freedom.  
Then each degree of freedom appended wipes out not only itself but also one of $\FrQ$'s redundancies, so one indeed ends up on the configuration space that is free of these redundancies. 
I.e. on the quotient space $\FrQ/\FrG$, as is required to successfully implement Configurational Relationalism.

In the Best Matching procedure, one furthermore takes $\scS\scH\scU\scF\scF\scL\scE_{\sfG}$ as equations in the Lagrangian variables to solve for the $g^{\sfG}$ auxiliaries themselves.
One then substitutes the extremizing solution back into the original action to obtain a reduced action on the $\FrQ/\FrG$ configuration space.

Let us further clarify this with the example of scaled RPM.  
These are taken to be fundamental rather than effective mechanics problems for which potentials of the form\footnote{In this paper, 
I use underline for spatial 3-vectors, overhead arrows for spacetime 4-vectors and bold font for configuration space quantities (and their conjugate momenta).} 
$V(\mbox{\boldmath{$q$}}) = V(\underline{q}_I \cdot \underline{q}_J  \mbox{ alone})$ apply.
This form then guarantees that auxiliary translation and rotation corrections applied to this part of the action straightforwardly cancel each other out within.  
[Because of this, there is no need in this example to correct the $\mbox{\boldmath{$q$}}$ themselves.]  
The situation with the kinetic term is more complicated because $\d/\d\lambda$ is not a tensorial operation under $\lambda$-dependent translations and rotations. 
This leads to the translation and rotation corrected kinetic term,   

\ni\be
T^{\sR\sP\sM} = ||\circ_{\underline{A}, \underline{B}}\mbox{\boldmath$q$}||_{\mbox{\scriptsize\boldmath$m$}}\mbox{}^2/2 
\mbox{ } \mbox{ for } \mbox{ } \circ_{\underline{A}, \underline{B}}\mbox{\boldmath$q$} := \dot{\mbox{\boldmath$q$}} - \underline{A} - \underline{B} \cr \mbox{\boldmath$q$} \mbox{ } .    
\label{T}
\ee
The action is then

\ni\be
\FS^{\sR\sP\sM} = 2 \int \d\lambda \sqrt{WT^{\sR\sP\sM}}  \mbox{ } .
\label{Jac} 
\ee
Then variation with respect to $\underline{A}$ and $\underline{B}$ give the secondary constraints

\ni\be
\underline{\scP} := \sum\mbox{}_{\mbox{}_{\mbox{\scriptsize I = 1}}}^{N} \underline{p}_{I} = 0 \mbox{ } , \mbox{ } \mbox{ }
\underline{\scL} := \sum\mbox{}_{\mbox{}_{\mbox{\scriptsize I = 1}}}^{N} \underline{q}^{I} \cr \underline{p}_{I} = 0 \mbox{ }  .
\label{ZM-AM}
\ee
These are, respectively, zero total momentum and angular momentum constraints, all of which are linear in the momenta. 
The first can furthermore be interpreted as the centre of mass motion for the dynamics of the whole universe being irrelevant rather than physical.  
All the tangible physics is in the remaining relative vectors between particles.

Let us now return to tackling the PoT in an ordered sequence.

\ni Step --2) As per the start of this Section, one can develop Configuration Relationalism resolution separately from Temporal Relationalism resolution, 
in terms of Lagrange Multiplier auxiliaries. 
However these then ruin the Manifest Reparametrization Invariance -- and upgrades -- that implement Temporal Relationalism: a first interference between these facets.

\ni Step --1) However, we have already explained the answer to this incompatibility: work ab initio within a TRiPoD formulation of Physics. 
To get               Manifest Reparametrization Invariance, use not a Lagrange multiplier coordinate but a cyclic velocity $\dot{g}$; 
in upgrading this to Manifest   Parametrization Irrelevance and then the geometrical form that happens to be dual to this, use a cyclic differential $\d g$.
Then indeed Relationalism points to the accompanying variation being free-end point\foo{Or end-spatial hypersurface in the case of geometrodyamics.} 
value variation, the outcome of which coincides with the conventional multiplier formulation's.   

\ni Step 0) The Configurational Relationalism nontrivial case's Temporal Relationalism resolving step has an extra element. 
I.e. rearranging $\scC\scH\scR\scO\scN\scO\scS$ involves its Lagrangian-variables form, 
but now this is an expression in terms of $\FrG$-auxiliaries which cannot have physical meaning attributed to them. 
This is resolved by extremizing over $\FrG$ in the particular form

\ni\beq
t^{\se\sm(\sJ)} = \mE_{\sg \in \FrG} \int \d s_{\sg}/\sqrt{2W}  \mbox{ } 
\label{t-em-J-G}
\eeq
this extremization (denoted by $\mE$) acting so as to wipe out the $\FrG$-dependence.
Furthermore, this extremization is of a second functional -- the action itself -- rather than of the obvious first functional in (\ref{t-em-J-G}); 
in general extremizing over the first functional itself is inconsistent \cite{FileR}.  
{\sl This forces one to tackle Configurational Relationalism prior to Temporal Relationalism}, hence explaining why the above steps have enumerations prior to the previous Sec's.

In the case of RPM, the Temporal {\sl and} Configurational Relationalism implementing action is 

\ni\beq
\FS^{\sR\sP\sM} = \sqrt{2} \int \d\lambda \sqrt{W}\d s^{\sR\sP\sM} \mbox{ } , \mbox{ } \mbox{ }
\d s^{\sR\sP\sM\, 2} := ||\d_{\underline{A}, \underline{B}}\mbox{\boldmath$q$}||_{\mbox{\scriptsize\boldmath$m$}}\mbox{}^2 
\mbox{ } \mbox{ for } \mbox{ } \d_{\underline{A}, \underline{B}}\mbox{\boldmath$q$} := \d{\mbox{\boldmath$q$}} - \d\underline{A} - {\d\underline{B}} \cr \mbox{\boldmath$q$} \mbox{ } .    
\label{Jac-2} 
\eeq
\mbox{ } \mbox{ } In the GR case, the analogue of (\ref{Jac}) is the BSW action, and that of (\ref{Jac-2}) the fully relational action 

\ni\be
\FS^{\sG\sR}_{\sr\se\sll} = \int\d\lambda\int_{\Sigma}\sqrt{\mh}\sqrt{\mR - 2\Lambda}\,\d \ms^{\sG\sR}_{\sr\se\sll}  \mbox{ } , 
\d s^{\sG\sR}_{\textrm{rel}} := \left|\left|\d_{\stackrel{\rightarrow}{\sF}}\bh\right|\right|_{\sbM}^{\mbox{ }\mbox{ } 2} \mbox{ } . 
\label{S-BSW}
\ee
Here $\mM^{abcd} := \sqrt{\mh}\{\mh^{ac}\mh^{bd} - \mh^{ab}\mh^{cd}\}$ is the GR configuration space metric, and $\d_{\sF}\mh_{ij} := \d \mh_{ij} - \pounds_{\sd \sF} \mh_{ij}$. 
This can be conceived of from Temporal and Configurational Relationalism's first principles rather than ever passing through ADM's formulation \cite{RWR, FileR, AM13}.

The primary constraint arising in this case is indeed the usual $\scH$.
The secondary constraint is $\scM_i$. 
The geometrodynamical subcase of Best Matching is the so-called {\it Thin Sandwich}: Fig \ref{Facet-Intro-4}.b) and \cite{BSW, WheelerGRT}. 
Originally, this involved solving the Lag variables form of $\scM_i$ for $\beta_i$.
In TRiPoD form it involves solving Jacobi--Mach variables form of $\scM_i$ for $\d \mF_i$, though this change makes no difference to the mathematical form of the problem.

To understand the name, consider first the {\it thick sandwich} prescribes knowns $\mh_{ij}^{(1)}$ and $\mh_{ij}^{(2)}$ on two hypersurfaces -- the `slices of bread' -- 
and one is to solve for the finite region of `filling' in between (Fig 1.b), in analogy with the QM set-up of transition amplitudes between states at two different times \cite{WheelerGRT}.
This turns out to be very ill-defined mathematically.
The thin sandwich is then Wheeler's \cite{WheelerGRT} `thin limit' of this, 
with spatial metric $\mh_{ij}$ and its label-time velocity $\dot{\mh}_{ij}$ prescribed as data on a spatial hypersurface $\Sigma$ (Fig 1.c).
Upon solving for $\upbeta^i$ or $\d \mF^i$, one constructs an infinitesimal piece of spacetime to the future of $\Sigma$ via forming the extrinsic curvature combination.  
This is another of Isham and Kucha\v{r}'s listed PoT facets (presented as a manifestly classical problem).   
The Thin Sandwich Problem remains a problem because its partial differential equation mathematics is hard \cite{TSC2}.  
That it is indeed a major mathematical problem is clear from e.g. \cite{TSC2}.

{            \begin{figure}[ht]
\centering
\includegraphics[width=0.75\textwidth]{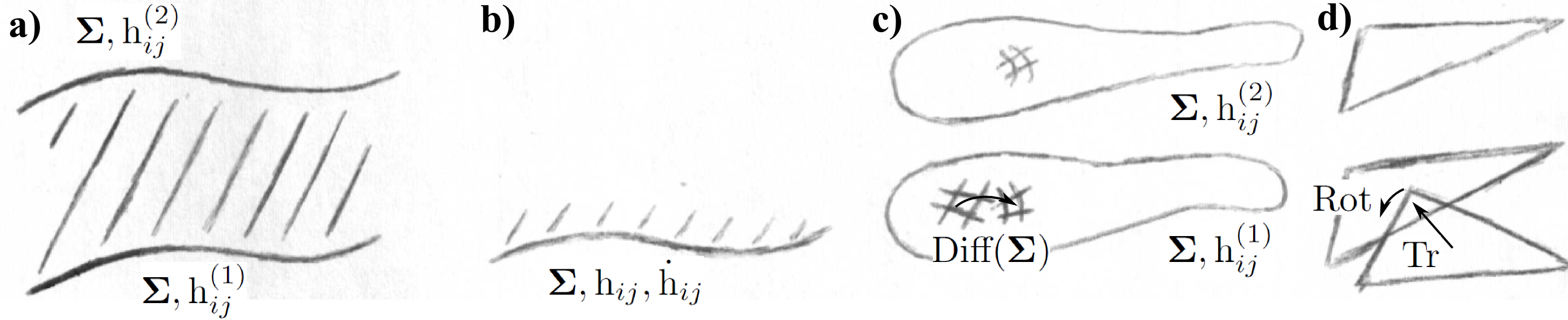}
\caption[Text der im Bilderverzeichnis auftaucht]{        \footnotesize{
a) Thick sandwich bounding bread-slice data to solve for the spacetime `filling'. 
b) Thin Sandwich data to solve for a local coating of spacetime \cite{WheelerGRT}.
This is the `thin' limit of taking the bounding `slices of bread': the hypersurfaces $\mh_{ij}^{(1)}$ and $\mh_{ij}^{(2)}$ as knowns.
This solved for the spacetime `filling' in between, in analogy with the QM set-up of transition amplitudes between states at two different times \cite{WheelerGRT}.
c) The Thin Sandwich can then be re-interpreted in terms of Best Matching Riem($\bupSigma$) with respect to Diff($\bupSigma$), which amounts to the shuffling as depicted.
I.e. keep one fixed  and shuffle the other to seek out how to minimize the incongruence between the two.  
d) The analogous triangleland Best Matching for relational triangles with respect to the rotations Rot and translations Tr.
This is included to indicate that Best Matching can be applied to a wide range of theories rather than just to geometrodynamics or the corresponding spatial diffeomorphisms.} }
\label{Facet-Intro-4} \end{figure}          }

\ni On the other hand, the RPM counterpart of the problem is resolved for 1- and 2-$d$ RPM's \cite{Kendall, FileR}.  
The constraints (\ref{ZM-AM}), rewritten in `Lagrangian' configuration--velocity variables 
($\mbox{\boldmath{$q$}}, \dot{\mbox{\boldmath{$q$}}}$), can here be solved for the auxiliary variables $\d\underline{A}$, $\d\underline{B}$ themselves.
This solution is then to be substituted back into the action, so as to produce a final Tr and Rot-independent expression that {\sl directly} implements Configurational Relationalism.  
This action is to be elevated this new action to be one's primary starting point.

Dealing with a Thin Sandwich Problem is unnecessary for minisuperspace \cite{AMSS1}, since here spatial homogeneity precludes nontrivial action of spatial diffeomorphisms.
It is resolved to leading order for inhomogeneous perturbations about isotropic spatially $\mathbb{S}^3$ minisuperspace with scalar field matter \cite{SIC1} 
(using a Machianized version of Halliwell--Hawking's model \cite{HallHaw}).

\mbox{ } 

\ni Finally, let us generalize the indirect implementation of Configurational Relationalism to the following `$\FrG$-act $\FrG$-all' method. 
Given an object $O$ that corresponds to the theory with configuration space $\FrQ$, one first applies a group action of $\FrG$ to this --- denoted $\stackrel{\rightarrow}{\FrG}_gO$. 
This amounts to making a $\FrG$-bundle version of $O$. 
Secondly, one applies some operation $\mbox{\Large S}_g$ that makes use of all of the $g^{\sfG} \in \FrG$ so as to cancel out the appearance of $g^{\sfG}$ in the group action.

Examples of $\mbox{\Large S}_g$ include summing, integrating, averaging, taking infs and sups, and extremizing, in each case over $\FrG$.

As a further generalization \cite{FileR, BI, ABook}, one can insert `Maps' in between the $\FrG$-act and $\FrG$-all moves to make a general metric background invariantizing (MBI) map

\ni\beq
\mbox{MBI} : \mbox{Maps} \circ O \mapsto O_{\sFG} = \mbox{\Large S}_{g \in \sFG} \circ \mbox{Maps } \circ \stackrel{\rightarrow}{\FrG} O \mbox{ } . 
\eeq
Best Matching is then a Lagrangian-level case with $O$ an action and using an infinitesimal $\FrG$ action and extremization over $\FrG$ for its $\FrG$-all operation
However, this implementation also allows Configurational Relationalism to be applied i) at levels other than that of the Lagrangian variables, 
such as at the Hamiltonian level or at the level of solving the quantum equations. 
ii) To objects entirely unlike actions, e.g. the finite-rotation shape comparer \cite{Kendall}, 
the Gromov--Hausdorff distance between metric spaces and the even more well-known {\it group averaging} technique from Group Theory and Representation Theory. 
The $\FrG$-all operations for these are extremize, inf and average respectively.  
Further examples are $\FrG$ invariant notions of distance more generally, of information, of correlation and of QM operators.
It is via this observation that Configurational Relationalism can be considered in the context of whatever other structures one's theorizing requires, 
in particular leading to its incorporation in the treatment of the other PoT facets .

\subsection{Constraint Closure (Problem)}\label{CC}

%
Do constraints beget more constraints? 
I.e. if $\scC_{\sfA}$ vanishes on a given spatial hypersurface, what can be said about $\dot{\scC}_{\sfA}$?  
If $\dot{\cal C}_{\sfA}$ is equal to some $f({\cal C}_{\sfA})$ alone, it is said to be {\it weakly zero} in the sense of Dirac \cite{Dirac} (denoted by $\approx 0$). 
However, there is a lack of rigour in such a `Lagrangian' picture of `constraint propagation' (i.e. $\dot{\scC}_{\sfA}$ evaluated by use of Euler--Lagrange equations).  
Let us next consider the joint space of $Q^{\sfA}$ and $P_{\sfA}$, 
as standardly equipped with the Poisson brackets structure $\mbox{\bf \{} \mbox{ } \mbox{\bf ,} \mbox{ } \mbox{\bf \}}$: {\it phase space}.
In this formulation, forming and handling brackets between constraints turns out to give a rigorous algorithm for handling whether constraints beget further independent constraints.
This is {\it Dirac's algorithm} \cite{Dirac, HTbook}, in which the brackets of known constraints can in general provide further constraints, specifier equations and inconsistencies.
This makes it clear that Constraint Closure is indeed a necessary check.
The end-product algebraic structure of constraints is, schematically, 

\ni\beq
\mbox{\bf \{} {\scC}_{\sfF}\mbox{\bf ,} \,  {\scC}_{\sfF^{\prime}} \mbox{\bf \}} \approx 0 \mbox{ } .
\label{C-C}
\eeq
$\fF$ here indexes {\it first-class} constraints: those that close among themselves under Poisson brackets.
A constraint is {\it second-class} if it is not first-class; 
first-class constraints use up 2 degrees of freedom each to second-class's 1; 
gauge constraints are a subset of first-class constraints.
See e.g. \cite{HTbook} for means of removing second-class constraints.

A Dirac-type algorithm can then be applied to determine whether the whole picture of a theory is afforded constraints provided by Temporal and Configurational Relationalism 
-- $\scC\scH\scR\scO\scN\scO\scS$ and $\scS\scH\scU\scF\scF\scL\scE_{\sfI}$ respectively, which are in particular $\scH$ and $\scM_i$ in the case of GR. 
[These would not be the whole picture if {\sl further} constraints or specifier equations arise upon forming the brackets among these provided constraints, or if these 
provided constraints do not behave amongst themselves in the expected manner.]    
First in particular, suppose the $\scS\scH\scU\scF\scF\scL\scE_{\sfI}$ close among themselves in the form 

\ni\beq
\mbox{\bf \{} \scS\scH\scU\scF\scF\scL\scE_{\sfI} \mbox{\bf ,} \, \scS\scH\scU\scF\scF\scL\scE_{\sfI^{\prime}} \mbox{\bf \}} = 
                                       f_{\sfI\sfI^{\prime}}\mbox{}^{\sfI^{\prime\prime}} \scS\scH\scU\scF\scF\scL\scE_{\sfI^{\prime\prime}}                  \mbox{ } , 
\eeq
where $f_{\sfI\sfI^{\prime}}\mbox{}^{\sfI^{\prime\prime}}$ are constants, so this is a Lie algebra.
Then the $\scS\scH\scU\scF\scF\scL\scE_{\sfI}$ arising from the attempted rendering irrelevant of $\FrG$ is vindicated, 
insofar as it is playing the role of a gauge algebra $\scG\scA\scU\scG\scE_{\sfG}$ that realizes the physical irrelevance of $\FrG$.

Second, consider whether  $\scC\scH\scR\scO\scN\scO\scS$ closes by itself.
For RPM it does and for GR it does not in general.

Thirdly, consider whether $\scS\scH\scU\scF\scF\scL\scE_{\sfI}$ and $\scC\scH\scR\scO\scN\scO\scS$ close between them. 
That involves both of the above brackets (now allowing for each bracket to produce a linear combination of both of the subdivisions of the constraints).  
For GR this is OK, since the bracket of two $\scC\scH\scR\scO\scN\scO\scS$' just produces a $\scS\scH\scU\scF\scF\scL\scE_{\sfI}$.
This last closure clearly also involves computing the cross-bracket 

\ni\beq
\mbox{\bf \{} \scC\scH\scR\scO\scN\scO\scS \mbox{\bf ,} \, \scS\scH\scU\scF\scF\scL\scE_{\sfI}\mbox{\bf \}} \mbox{ } .
\eeq
If this closes by producing just $\scC\scH\scR\scO\scN\scO\scS$, $\scC\scH\scR\scO\scN\scO\scS$ is established as a good $\FrG$-object: scalar densities in the RPM and GR cases.  
Were these brackets to produce new \scF\scL\scI\scN, however, they  would be asserting that an enlarged $\FrG$ is a necessary consideration.

Moreover, TRiPoD requires a variant of the above, involving differential almost phase space, differential almost-Hamiltonian 
and an almost-Dirac procedure, in which the appending is with cyclic differentials. 
Most of this distinction is in case for which the $\FrG$ auxiliaries can not be eliminated.

To present the GR case, first note that field theory constraint algebras are most usefully presented in terms of smearing functions,\footnote{In general, I use 
$({\scC}_{\tfW}| \sA^{\tfW}) := \int d^3z \, {\scC}_{\tfW}(z^i; \upphi_{\tfA}] \sA^{\tfW}(z^i)$ (an `inner product' notation) 
for the {\it smearing} of a $\fW$-tensor density valued constraint ${\cal C}_{\tfW}$ by an opposite-rank $\fW$-tensor smearing with no density weighting: $A^{\tfW}$. 
$\upphi_{\tfA}$ is here a full repertoire of the fields involved.
The undefined letters in this Sec's presentation of GR are smearings of this kind.}  
%
the presentation of which is slightly reformulated to lie within TRiPoD. 
Then using $X \overleftrightarrow{\pa}^i Y := \{ \pa^i Y \} X - Y \pa^i X$, 

\ni \be
\mbox{\bf \{} (    \scM_i  |    \upxi^i    ) \mbox{\bf ,} \, (    \scM_j    |    \d \upchi^j    ) \mbox{\bf \}} =  (    \scM_i    | \, [ \upxi, \upchi ]^i )  \mbox{ } ,
\label{Mom,Mom}
\ee

\ni \be
\mbox{\bf \{} (    \scH    |    \upzeta    ) \mbox{\bf ,} \, (    \scM_i    |    \upxi^i    ) \mbox{\bf \}} = (    \pounds_{\underline{\upxi}} \scH    |    \upzeta    )  \mbox{ } , 
\label{Ham,Mom}
\ee

\ni\be 
\mbox{\bf \{} (    \scH    |    \upzeta    ) \mbox{\bf ,} \,(  \scH  |    \upomega  )\mbox{\bf \}}  = (  \scM_i \mh^{ij}   |   \upzeta \, \overleftrightarrow{\pa}_j \upomega )  \mbox{ } . 
\label{Ham,Ham}
\ee
Note that this does close in the sense that there are no further constraints or other conditions arising in the right-hand-side expressions. 
The first bracket means that Diff($\bupSigma$) on a given spatial hypersurface themselves close as an (infinite-dimensional) Lie algebra.
The second bracket means that $\scH$ is a good object (a scalar density) under Diff($\bupSigma$).  
Both of the above are kinematical rather than dynamical results.    
The third bracket, however, is more complicated in both form and meaning \cite{T73}.

Note that the presence of $\mh^{ij}(\mh_{kl}(\underline{x}))$ in its right hand side expression causes 
the transformation itself to depend upon the object acted upon -- contrast with the familiar case of the rotations \cite{T73}!

Also note that the GR constraints to form a more general algebraic entity than a Lie algebra: a {\it Lie algebroid}.   
More specifically, (\ref{Mom,Mom}--\ref{Ham,Ham}) form the {\it Dirac algebroid} \cite{Dirac}.     
(See \cite{BojoBook} if interested in a bit more on algebroids in general and the Dirac algebroid in particular.)

Next, if one tried to do GR without Diff($\bupSigma$) irrelevance, (\ref{Ham,Ham}) would in any case enforce this.

Finally, by not forming a Lie algebra, clearly the constraints -- and Diff($\FrM$, Fol) -- form something other than Diff($\FrM$).  
Indeed, the vast difference in size corresponds to the variety of possible foliations.

At the quantum level for field theories, the Constraint Closure Problem becomes the Functional Evolution Problem -- a listed PoT facet.
This might occur even if classical Constraint Closure is guaranteed, since

\ni \be
\widehat{\cal C}_{\sfC} \Psi = 0 \mbox{ } \not{\Rightarrow \mbox{}} \mbox{ } \mbox{\bf [}\widehat{\cal C}_{\sfC}\mbox{\bf ,} \, \widehat{\cal C}_{\sfC^{\prime}}\mbox{\bf ]}\Psi = 0 
\mbox{ } . 
\label{FEP}
\ee 
Instead, more constraint terms might be unveiled, or the right-hand-side of the second equation in (\ref{FEP}) might not be zero but rather contain an anomaly term.  
For GR in general, this remains an unsolved problem.

\subsection{Expression in Terms of Beables, and the Problem of Beables}

Having found constraints and introduced a classical brackets structure, one can then ask which objects have zero classical brackets with `the constraints'.
These objects, termed {\it observables} or {\it beables}, are more physically useful than just any $Q^{\sfA}$ and $P_{\sfA}$ due to containing physical information only.  
The Jacobi identity applied to two constraints and one observable/beable determines that \cite{ABeables}
\beq
\mbox{the input notion of `the constraints' is a closed algebraic structure of constraints} \mbox{ } .
\label{CCB}
\eeq
Applied instead to one constraint and two observables/beables, it determines that the observables/beables themselves form a closed algebraic structure.  
In this sense, the observables/beables form an algebraic structure that is associated with that formed by the constraints themselves.

Also note the contextual distinction between observables, which `are observed', and beables, which just `are'. 
Bell \cite{Bell} then argued that the latter are more appropriate for whole-universe cosmology, 
via carrying no connotations of external observing; there is further quantum-level motivation for beables; see also \cite{ABeables}.

In particular, {\it Dirac beables} are quantities that (for now classical) brackets-commute with all of a given theory's first-class constraints, 

\ni\beq
\mbox{\bf \{}    \scC_{\sfC}    \mbox{\bf ,} \, \iD_{\sfD}   \mbox{\bf \}} \mbox{ } `=' 0 \mbox{ } .
\eeq 
Examples include $\scE$, $\scL_i$, $\scP_i$ for RPM, or with $\scH$, $\scM_i$ for GR.

On the other hand, {\it \K beables} \cite{Kuchar93} are quantities that form zero brackets with all of a given theory's first-class linear constraints, 

\ni\beq
\mbox{\bf \{}    \scF\scL\scI\scN_{\sfI}    \mbox{\bf ,}  \, \iK_{\sfK}    \mbox{\bf \}} \mbox{ } `=' 0 \mbox{ } .
\eeq
In common examples, $\scF\scL\scI\scN_{\sfG} = \scG\scA\scU\scG\scE_{\sfG}$ (see \cite{HTbook} for a counter-example), 
in which case \K beables coincide with {\it gauge-invariant quantities}. 
The gauge-invariant quantities in question correspond to the successful realization of Configurational Relationalism's $\FrG$ 
as a means of passage to a less redundant quotient configuration space $\FrQ/\FrG$.  
As examples, \K beables are trivial for minisuperspace and spatially-absolute Mechanics, RPM examples include pure shapes and scales.
There is no Dirac--Kucha\v{r} beables distinction for Electromagnetism, and \K beables for GR include, formally, the 3-geometries themselves.

Finally, one can consider the complementary notion of quantities such that just 

\ni\beq
\mbox{\bf \{}  \scC\scH\scR\scO\scN\scO\scS \mbox{\bf ,}  \,  \iB_{\sfB} \mbox{\bf \}} \mbox{ } `=' 0 \mbox{ } . 
\eeq    
[These are just Dirac beables in the absense of $\scF\scL\scI\scN_{\sfG}$, e.g. for minisuperspace.]

These can be viewed \cite{APoT3} as separate aspects (subaspects of the notion of expression in terms of beables), 
corresponding to algebraic structures associated with each type of constraint.
Not all are meaningfully realized, need underlying algebraic closure of constraints used in order to meaningfully posit a type of beable.

Finally, the {\it Problem of Beables} -- more usually termed `Problem of Observables': Facet 4) of the PoT -- 
is that it is hard to construct a set of beables, in particular for gravitational theory.  
This is problematic because \K and especially Dirac beables are hard to find \cite{ABeables}, 
and even more so at the quantum level (where the classical brackets are replaced with quantum commutators).
Finally, 

\ni\beq
\mbox{\bf \{}  \scC_{\sfC} \mbox{\bf ,}  \,  \iD_{\sfD} \mbox{\bf \}}\mbox{ } `=' 0 \mbox{ } \not{\rightarrow} 
\mbox{\bf  [}  \scC_{\sfC} \mbox{\bf ,}  \,  \iD_{\sfD} \mbox{\bf  ]} \mbox{ } `=' 0
\eeq 
and similarly for other types of beables.

\subsection{Spacetime Relationalism}

GR has more background independent features than theories of Mechanics do.
This is due to GR having a spacetime notion, which has more geometrical content than Mechanics' space-time notion does.
The latter is far more of an amalgamation of separate space and time notions 
-- multiple copies of a spatial geometry strung together by being labelled with a time variable -- whereas the latter is a co-geometrization of space and time. 
In the GR case, spacetime possesses its own versions of generator-providing Relationalism, closure of these generators and beables as associated commutants with these generators.

\mbox{ }

\ni This subsection's position is for now to start afresh with primality now ascribed to spacetime rather than to space.  
Spacetime's own Relationalism is then as follows. 

\ni i) The are to be no extraneous spacetime structures, in particular no indefinite background spacetime metrics. 
Fixed background spacetime metrics are also more well-known than fixed background space metrics. 

\ni ii) Now as well as considering a spacetime manifold $\FrM$, consider also a $\FrG_{\sS}$ of transformations acting upon $\FrM$ that are taken to be physically redundant.

\ni For GR,  $\FrG_{\sS}$ = Diff($\FrM$).
Also note that $\FrM$ can be equipped with matter fields in addition to the metric. 
Then i) can be extended to include no extraneous internal structures, 
whereas ii)'s $\FrG_{\sS}$ can have a part acting internally on a subset of the fields, corresponding to a spacetime/path format Gauge Theory.
Then the internal part of ii) is closer to QFT's more usual spacetime presentation of Gauge Theory than the internal part of Configurational Relationalism is.
On the other hand, Configurational Relationalism is more closely tied to Dirac observables/beables, out of these all being configuration-based notions.

\ni Aspect 5.b) Diff($\FrM$) indeed straightforwardly forms a Lie algebra, in parallel to how Diff($\bupSigma$) does:   

\ni \be
\mbox{\bf |[} (    \mD_{\mu}    |    X^{\mu})   \mbox{\bf ,} \, (    \mD_{\nu}|Y^{\nu}    ) \mbox{\bf ]|} \mbox{ } `=' (    \mD_{\gamma}    | \, [X, Y]^{\gamma}    ) \mbox{ } . 
\label{Lie-2} 
\ee
Diff($\FrM$) also shares further specific features with Diff($\bupSigma$), such as its right hand side being of Lie derivative form.
Thus all three types of Relationalism considered so far are implemented by Lie derivatives.

However, whereas Diff($\bupSigma$)'s generators are conventionally associated with dynamical constraints, Diff($\FrM$)'s are not. 
Additionally, Diff($\bupSigma$)'s but not Diff($\FrM$)'s classical realization of the Lie bracket is conventionally taken to be a Poisson bracket.
This furthermore implies that there is conventionally no complete spacetime analogue of the previous Sec's notion of beables/observables.
These differences are rooted in time being ascribed some further distinction in dynamical and then canonical formulations than in spacetime formulations.   
(\ref{Lie-2}) is to be additionally contrasted with the Dirac algebroid (\ref{Mom,Mom}--\ref{Ham,Ham}).  
Clearly there are two very different algebraic structures that can be associated with GR spacetime: 
the first with unsplit spacetime and the second with split space-time including keeping track of how it is split.

For further detail of background independent concepts and terminology, consult \cite{A6467Giu06, APoT3}.

I also emphasize GR spacetime's standard modelling of multiplicity of observers and of causality. 
This carries over to other theories which presuppose such a spacetime structure, for instance higher curvature theories and scalar--tensor theories.

\ni Aspect 5.c) Spacetime Observables.
Diff($\FrM$) is closely related to {\it spacetime observables} in GR.
Such objects would be manifestly Diff($\FrM$)-invariant. 
I.e. commutants $\mS_{\sfQ}$ [e.g. using the {\sl generator} version of weakly vanishing] 

\ni \be
\mbox{\bf |[} (    {\cal D}_{\mu}    |    X^{\mu})  \mbox{\bf ,} \, (    \mS_{\sfQ}|Y^{\sfQ}    ) \mbox{\bf ]|} \mbox{ }  `=' \mbox{ } 0 \mbox{ } \label{Sp-Obs} .
\ee
Note that this is tighter at the quantum level, e.g. Diff($\FrM$)-invariant measures would be needed for explicit realization of a quantum path integral approach.

\subsection{Foliation Independence, Foliation Dependence Problem, and Refoliation Invariance resolution.}    

Here one is to further develop embeddings, slices and foliations as more advanced foundations for the spacetime-assumed ADM split \cite{ADM}.
Moreover, GR spacetime admits multiple foliations.  
At least at first sight, this property is lost in the geometrodynamical formulation.

{\it Foliation Independence} is an aspect of Background Independence.
{\it Foliation Dependence} is then a type of privileged coordinate dependence. 
This runs against the basic principles of what GR contributes to Physics.
The {\it Foliation Dependence Problem} is the corresponding PoT facet. 
It is obviously a time problem since each foliation by spacelike hypersurfaces being orthogonal to a GR timefunction.
I.e. each slice corresponds to an instant of time for a cloud of observers distributed over the slice.
Each foliation corresponds to the cloud of observers moving in a particular way.

{\it Refoliation Invariance} is then that evolving via Fig \ref{Refol-5-SRPs}.e)'s dashed or dotted hypersurfaces gives the same physical answer \cite{T73}. 
This happens to be a property possessed by classical split GR space-time \cite{T73}.
This is by bracket (\ref{Ham,Ham})'s pictorial form telling one how to close Fig \ref{Refol-5-SRPs}.a) in the form of Fig \ref{Refol-5-SRPs}.b).  
Thus the Dirac algebroid not only guarantees constraint closure but also resolves the Foliation Dependence Problem at the classical level (Refoliation Invariance Theorem of GR).  
Encoding Diff($\FrM$, Fol) for arbitrary (rather than fixed) foliation Fol can now be seen as the reason why the much larger Dirac algebroid has replaced unsplit spacetime's Diff($\FrM$) 
[the Dirac Algebroid {\sl is} Diff($\FrM$, Fol)].

By the Refoliation Invariance property, GR spacetime is not just a strutting together of spaces like Newtonian space-time. 
Rather, it manages to be many such struttings at once that are physically mutually consistent.
Furthermore, this strutting is tied to the freedom of families of observers to move as they please.
Even the relational form of Mechanics fails to manifest this property; minisuperspace is often just considered in terms of the foliation privileged by spatially homogeneous surfaces. 
Both these limitations are unsurprising: Refoliation Invariance is a diffeomorphism-specific issue.

Finally, Hojman-Kucha\v{r}-Teitelboim \cite{HKT} obtained as a  first answer to Wheeler's question (\ref{Wheeler-Q}). 
Their first principles are the {\it deformation algebroid} of the two operations in Fig \ref{Refol-5-SRPs}c)-d) for a hypersurface, which take the same form as the Dirac algebroid.  
This set-up does however still presuppose spacetime (more specifically, embeddability into spacetime).

Note that most usually one makes a choice to work with split or unsplit spacetime.
However, a few Histories Theory approaches do involve both at once. 
Thus all of Temporal, Spatial and Spacetime Relationalism can be manifested at once.

See Fig \ref{Refol-5-SRPs}.e) for an outline of some of the consequences; \cite{TRiFol} considers these in more detail.  

{            \begin{figure}[ht]
\centering
\includegraphics[width=1.0\textwidth]{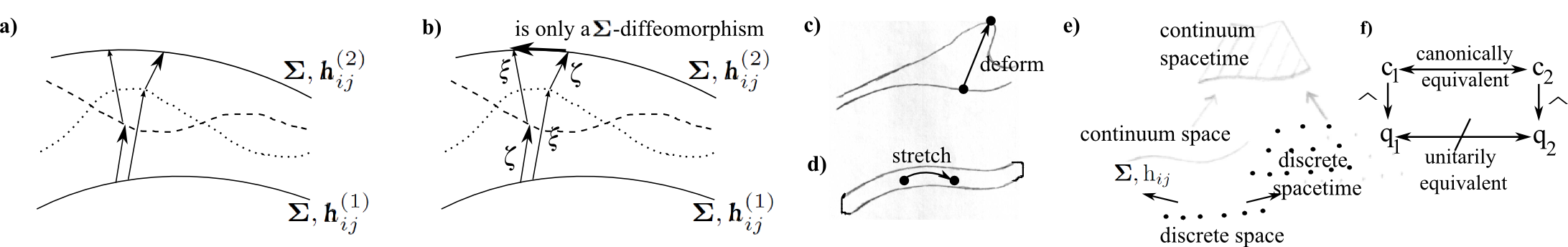}
\caption[Text der im Bilderverzeichnis auftaucht]{\footnotesize{a) The Foliation Dependence Problem is whether evolution via each of the dashed and the dotted spatial hypersurfaces 
give the same physical answers.  
b) Teitelboim's \cite{T73} classical `Refoliation Invariance' resolution of this via the pictorial form of the Dirac algebroid's bracket (\ref{Ham,Ham}). 
c) The pure deformation and d) the pure stretch (just the spatial diffeomorphism again).
e) Pictorial form of 3 types of Spacetime Construction: spacetime from space, from discrete spacetime and from discrete space.
f) supports the statement of the Multiple Choice Problem. 
Here `c' stands for classical formulation, `q' for quantum formulation and $\widehat{\mbox{ }}$ denotes quantization map.} }
\label{Refol-5-SRPs}\end{figure}            }

\subsection{Spacetime Construction (Problems)}\label{SCP}

Next consider assuming less structure than spacetime.  
In general, if classical spacetime is not assumed, one needs to be able to construct it, at least under suitable limiting conditions. 
This can be hard, particularly as the amount of structure assumed is lessened.

Already at the classical level, Spacetime Construction can be considered along two logically independent lines: 

\ni a) from space (as an `embed rather than project' `inverse problem' to the previous Sec's, which is harder since now only the structure of space is being assumed).

\ni b) From making less assumptions about continua, giving a total of four construction procedures: Fig \ref{Refol-5-SRPs}.e).  
I expand on this in the sense of `less layers of mathematical structure assumed' in Fig \ref{Bigger-Set-2}.

As Wheeler also pointed out \cite{Battelle}, at the quantum level, fluctuations of the dynamical entities are inevitable. 
In the present case, these are fluctuations of 3-geometry, and these are then too numerous to be embedded within a single spacetime.  
The beautiful geometrical way that classical GR manages to be Refoliation Invariant breaks down at the quantum level.

Wheeler furthermore pointed out \cite{Battelle} the further relevance of Heisenberg's uncertainty principle.  
Precisely-known position $\underline{q}$ and momentum $\underline{p}$ for a particle are a classical concept corresponding to a worldline.
This view of the world is entirely accepted to break down in quantum physics due to Heisenberg's Uncertainly Principle.
In QM, worldlines are replaced by the more diffuse notion of wavepackets. 
However, in GR, what the Heisenberg uncertainty principle now applies to are the quantum operator counterparts of $\mh_{ij}$ and $\mp^{ij}$. 
But by the well known relation between gravitational momentum and extrinsic curvature, this means that $\mh_{ij}$ and $\mK_{ij}$ are not precisely known.   
Thus the idea of embeddability of a 3-space with metric $\mh_{ij}$ within a spacetime is itself quantum-mechanically compromised.
Thus (something like) the geometrodynamical picture (considering the set of possible 3-geometries and the dynamics of these) 
would be expected to take over from the spacetime picture at the quantum level.  
It is then not clear what becomes of notions that are strongly associated with classical GR spacetime, 
such as locality (if one believes that the quantum replacement for spacetime is `foamy' \cite{Battelle}), or causality.   
In particular, microcausality is violated in some such approaches \cite{I81-I85, I93}.
Additionally, the recovery of semiclassicality aspect of spacetime construction has a long history of causing difficulties in Loop Quantum Gravity.  
The extension of such work to a family of such algebraic structures in parallel with the current paper remains to be tackled (this Sec is part-based on \cite{AM13}).

Finally note that for instance Newtonian Gravity and Strong Gravity do {\sl not} have a `standard GR type' spacetime structure \cite{Stewart, Henneaux79}.
This has implications as regards how observers and causality are modelled in GR largely not carrying over to these theories.

\subsection{Global Well-definednesses and Global Problems of Time}         

This is the final facet's renaming, emphasizing its even greater plurality: 
it can concern globality in space, time itself, spacetime, configuration space, phase space, classical solution space, Hilbert space, spaces of quantum operators...
Another classification of Global Problems of Time is into effects understandable in terms of meshing conditions of charts, of p.d.e. solutions, of representations or of unitary evolutions.
Some form or other of it affects almost all facets and strategies \cite{Kuchar92, I93, ABook}.

\subsection{Physically Accounted for Multiplicities and Multiple Choice Problems} 

This is only relevant once the quantum level is under consideration. 
As Fig 4.f) illustrates, it consists of canonical equivalence of classical formulations of a theory not implying unitary equivalence of the quantizations of each \cite{Gotay00}.  
By this, different choices of timefunction can lead to inequivalent quantum theories.
I rename this in the plural since it applies to choices of frame and of beables as well as to choices of time.

\subsection{Classifying facets}\label{7-Gates}

{            \begin{figure}[ht]
\centering
\includegraphics[width=0.45\textwidth]{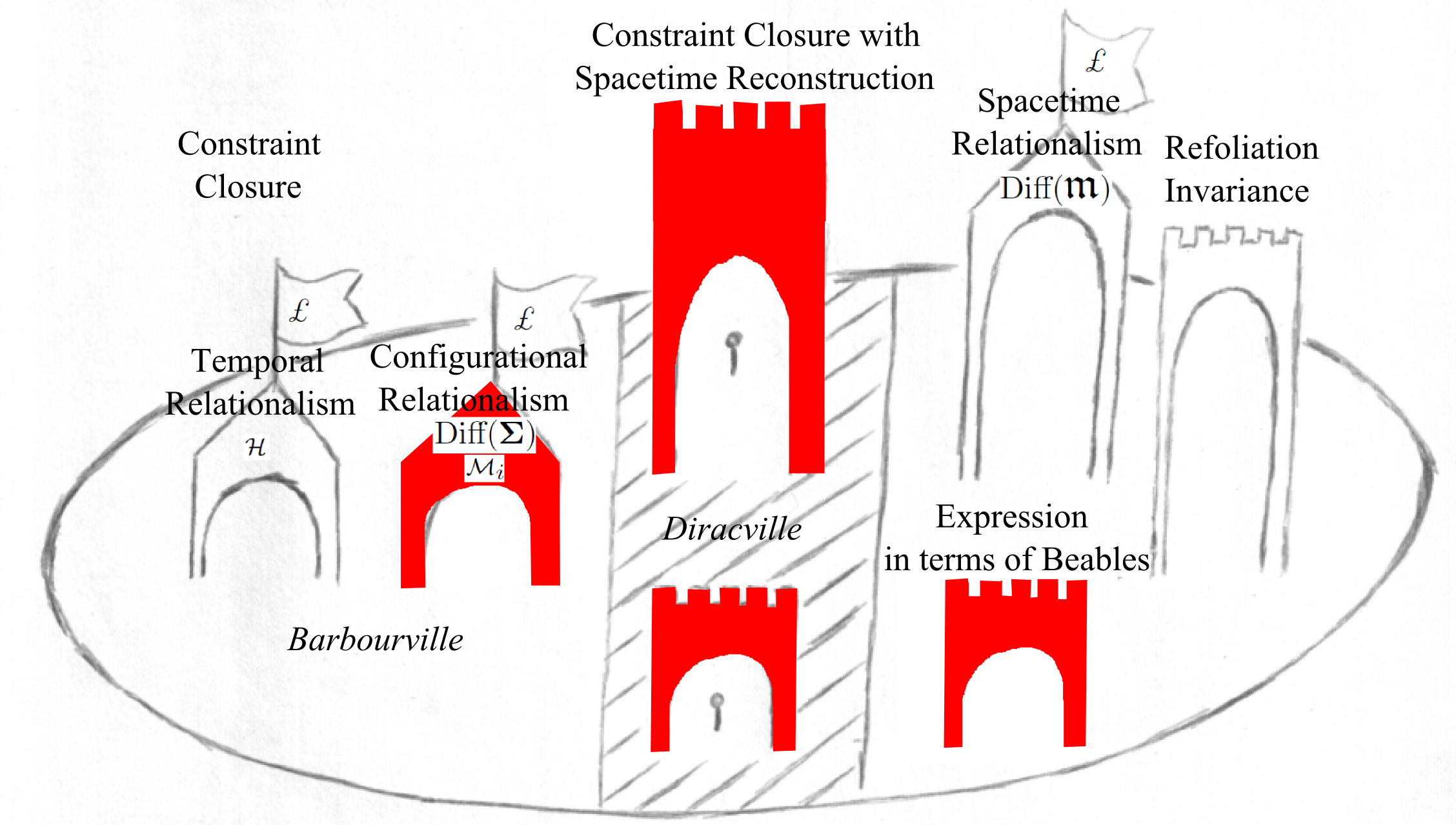}
\caption[Text der im Bilderverzeichnis auftaucht]{\footnotesize{Implementing each aspect of Background Independence causes a PoT facet to appear, so I depict these as gates.
For `a local' resolution of the PoT one needs to consistently get past seven of these (facets 1--7).
My presentation of this as a picture as well as a parable is by grouping and decorating the gates to indicate a number of significant subsets of the gates.
Groupings of gates indicated are, zonally, 
Barbourville and 
Diracville (also with keyholes indicated), 
spacetime (tall gates) versus space (short gates), and  
relational (pointy, and each admitting a Lie implementation as marked on its flag). 
All gates have an algebraic element to either their definition [Diff($\bupSigma$) and Diff($\FrM$) as indicated], or to their classical resolution (listed below).  
Some means through some gates at the classical level are as follows. 
Emergent Machian time to go through Temporal Relationalism (algebraically a reparametrization or a pure deformation) and Best Matching through Configurational Relationalism.
Nontrivial termination of the Dirac Procedure to unlock Constraint Closure, and likewise with Spacetime Construction for the taller double-gate version.
[This is algebraically the Dirac Algebroid = Diff($\FrM$, Fol) for the first and this singled out from a larger family of algebroids in the second; 
these are valid keys that open the keyholes].
Teitelboim's Refoliation Invariance depiction of the Dirac Algebroid = Diff($\FrM, \FrF$)'s $\{\scH, \scH\}$ bracket secures passage through Foliation Dependence. 
Note that the preceding trio are jointly resolved by the Dirac Algebroid in the case of classical GR.  
The Problem of Beables is to be resolved by finding an algebraic structure of beables associated with Dirac Algebroid = Diff($\FrM, \FrF$).
Finally, the four facets indicated in red correspond to algebraic structures involving Poisson Brackets.} } 
\label{Gates} \end{figure}          }
 
\ni Fig \ref{Gates} summarizes the present Sec's outline of each PoT facet as a gate, expanding on a quantum-level presentation of Kucha\v{r}'s \cite{Kuchar93}. 
{\sl There is a strong tendency for PoT facets to interfere with each other rather than standing as independent obstacles} \cite{Kuchar92, I93, Kuchar93}. 
The main point of the gates picture is that going through a further gate has a big tendency to leave one outside of gates that one had entered earlier.  
I.e. the facets bear rich conceptual and technical relations with each other since they arise from a joint cause: bridging the gap between background dependent and background independent 
paradigms in Physics, most notably the mismatch of the notions of time in GR and Quantum Theory.
Due to this, it is likely to be advantageous to treat them as parts of a coherent package rather than disassembling them into a mere list of problems to be addressed piecemeal.

Note the further multiplicity of some gates (see \cite{APoT3, ABook} for more).  
For instance, spacetime-related gates have histories counterparts. 
There are also multiple types of beables (Fig \ref{Bigger-Set-2}).
This further clarifies why Background Independence and the PoT have the list of constituent parts that they do.

\section{A local classical resolution}

\subsection{Temporal and Configurational Relationalism}

Begin by applying steps 1) to 4). 
If Configurational Relationalism is nontrivial, then apply steps --2) to 0) and then 1) to 4) again (the `loop the loop' in Fig \ref{Gates-2}.a).  

{            \begin{figure}[ht]
\centering
\includegraphics[width=1.0\textwidth]{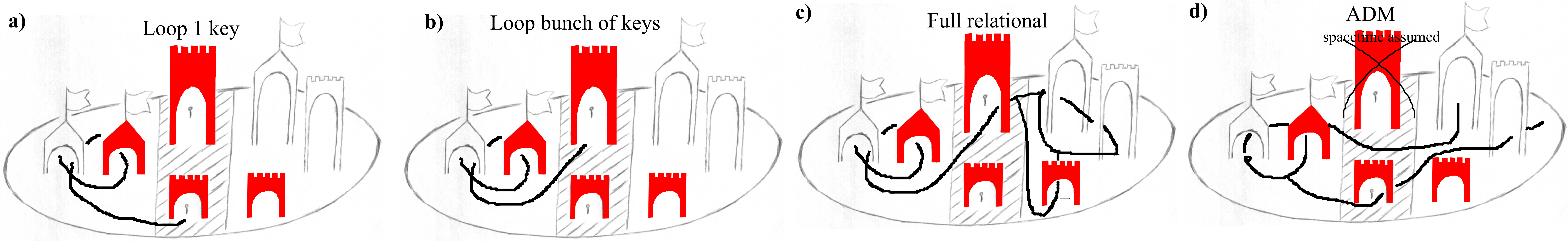}
\caption[Text der im Bilderverzeichnis auftaucht]{\footnotesize{a) The Configurational and Temporal Relationalism `loop the loop', 
producing a candidate key with which to open the Constraint Closure gate. 
b) Repeat this with a family of theories to produce a bunch of keys, one of which attains Spacetime Construction as well as constraint closure.
In trying each of a) and b), if one's $\FrQ$, $\FrG$, $\FS$ triple fails to produce any keys, go back to the loop and try again.
c) For contrast, the ADM and Teitelboim type alternative.
Here spacetime is assumed and implements 4-diffeomorphisms; in the Teitelboim version, the split of these with respect to a foliation is also considered. 
In any case, the ADM split then supplies $\scH$ and $\scM_i$ together.
Moreover, one could treat $\scM_i$ second here, since it is an integrability of $\scH$, ordering rather than splitting this path.
d) Back to b), having constructed spacetime, one can implement its relationalism and see how it splits (now in a Temporal Relationalism complying version). 
In parallel, and in either order with respect to the previous extension, one can have a go at constructing beables.} } 
\label{Gates-2} \end{figure}          }
 
\subsection{Constraint Closure}

\ni Step 8) Issue of whether the constraints provided are what we think.  
This is needed for the implementation of Configurational Relationalism to hold. 

\ni One can see this as whether a $\FrQ$, $\FrG$, $\FS$ triple stands up to Dirac's procedure... in fact almost-Dirac's procedure to remain within Temporal Relationalism.  
If $\scL\scI\scN$ is not first class, implementation of Configurational Relationalism is in question.  
If further constraints appear, see if they can be encoded, may pass one to a bigger $\FrG$.
In these cases, one is turned away at the third gate; one has to go back to the first two and try another $\FrQ$, $\FrG$, $\FS$ combination, 
until one has formed a successful key to the third gate.
There are two commuting orders through the other gates.

Note that if a given triple fails, one can restart with a new $\FrG$.  
In some cases the constraint closure produces a new constraint with specifically points to how to enlarge $\FrG$. 
This occurs for instance if one tries to set up GR with $\FrG$ = id rather than $\FrG$ = Diff($\bupSigma$). 
Then the momentum constraint arises as an integrability, thus pointing to the necessity of enlarging id to Diff($\bupSigma$).

Finally, some such changes of $\FrG$ furthermore transcend between levels of mathematical structure.
The above example passes from entirely involving Riemannian geometry to enforcing defferential geometry level Diff-invariance. 
Passing to conformal transformations and diffeomorphisms also corresponds to a particular level of mathematical structure (intermediate between the above two, 
as per Fig \ref{Bigger-Set-2}).
Finally, some alterations of $\FrG$ amount to descending to deeper levels of mathematical structure.

\subsection{Kucha\v{r} beables}

\ni Step 9) A second consequence \cite{AHall, FileR} of resolving Configurational Relationalism (the first being explicit $t^{\se\sm}$) 
is that one is in possession of a set of classical \K beables (a slight variant of the standard, to maintain Temporal Relationalism compatibility). 
For minisuperspace the concept is trivial, for RPM's, these are functions of shapes, scale and their conjugate momenta, 
and see \cite{SIC1} for the inhomogeneous perturbations about minisuperspace counterpart of these.
We'll return later to converting this to additionally resolving Dirac Beables.  
This structurally comes after Constraint Closure since it a consideration of associated algebraic structures.

\subsection{Extent of development of model arenas}

Whereas minisuperspace exhibits all but one of the aspects of Background Independence, I have explained how 
homogeneity  implies {\sl very quickly resolved} Problem of Time facets for many of these. 
On the other hand, RPM's exhibit 6 of the 9 traditional PoT aspects \cite{FileR}: see Fig \ref{Medium-Table}  
This includes the Configurational Relationalism aspect that minisuperspace does not possess. 

{            \begin{figure}[ht]\centering\includegraphics[width=1.0\textwidth]{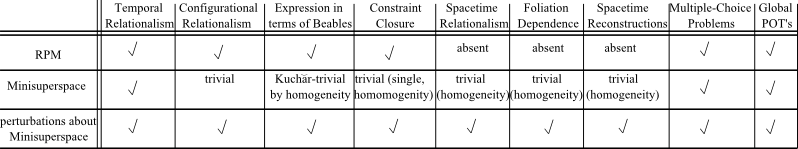}
\caption[Text der im Bilderverzeichnis auftaucht]{        \footnotesize{Which model arenas exhibit which facets.}  } \label{Medium-Table}\end{figure}          }

Moreover, while the RPM and minisuperspace cases are simple to calculate with, they miss the subtleties specifically associated with diffeomorphisms \cite{Kuchar92, I93}.  
A first arena in which these appear nontrivially is {\it slightly inhomogenous cosmology}.  
A particular such involves inhomogeneous perturbations about the spatially-$\mathbb{S}^3$ minisuperspace with single scalar field matter model.  
At the semiclassical quantum cosmology level, this particular case becomes the {\it Halliwell--Hawking model} \cite{HallHaw, SIC1}, 
which I choose as the current Seminar's most complicated specific example.    
RPM's and minisuperspace complementarily support by one or the other having all {\sl other} Background Independence aspects and consequent Problem of Time factes of this model.   
Here one considers the first few (usually two) orders of the perturbation of the metric.
Each of these form a simplified configuration space in place of the full Riem($\bupSigma$) that are currently under investigation.
\cite{SIC1} shows that slightly inhomogeneous cosmology already exhibits the Thin Sandwich problem, which in this case, moreover, is solvable, and also considers  
this model arena from a Histories Theory perspective.

\subsection{Spacetime Construction}\label{SCP-2}

\ni Step 10) One can also approach Constraint Closure for families of theories \cite{RWR, AM13}. 
This can give {\sl a bunch} of valid keys.

\ni Step 11) A further possibility is that Dirac procedure serves to construct spacetime.
Instead of assuming the GR form of the kinetic arc element and potential (given by presupposing spacetime and splitting it along the lines of ADM), consider the 4-parameter family ansatz 

\ni \be
S^{w,y,a,b} = \int \int_{\Sigma} \d^3x \sqrt{\sqrt{\mh}\{a \, \mR + b\}} \, \d \ms_{w,y} \mbox{ } . 
\label{trial} 
\ee
Here, $\d \ms_{w,y}$ is built out of the usual $\d_{\underline{\sF}}$ 
and the more general if still ultralocal supermetric $\bM_{w,y}$ with components $\mM^{abcd}_{w,y} := \sqrt{\mh}\{\mh^{ac}\mh^{bd} - w \, \mh^{ab}\mh^{cd}\}/y$.  
Then this leads to a four-pronged fork: 4-factor obstruction term \cite{AM13} 

\ni \be
2 \times a \times y \times \{ 1 - x \} \times (    \mD_i \mp \, | \, \pa \mJ \, \overleftrightarrow{\pa}^i \pa \mK    ) \mbox{ }   
\label{4-Factors}
\ee 
from the `smeared out' version of $\mbox{\bf \{} \scH \mbox{\bf ,} \, \scH \mbox{\bf \}}$.
The third prong \cite{RWR} amounts to a recovery of GR's Hamiltonian constraint. 
This is then part of the embeddability condition into a surrounding spacetime.  
Upon inclusion of simple fundamental matter fields, this comes hand in hand with a local Lorentzian relativity
The second prong gives an in general curved geometrostatics with Galilean relativity.
Thus these two factors represent Einstein's historical dichotomy, now arising from the Dirac constraint procedure!
Actually, it is a trichotomy, since zero as well as infinite propagation speed is a logical possibility: a distinct simpler geometrodynamics than GR's with Carrollian relativity.  
The first prong is of this nature.

The fourth prong -- a weakly vanishing option unlike the other three strongly-vanishing ones -- gives the CMC condition $\mp/\sqrt{\mh}$ = const \cite{RWR}.
Following this up \cite{ABFKO} by encoding the corresponding volume-preserving conformal transformations VPConf($\bupSigma$) is one path to shape dynamics \cite{Kos2, Mercati14}.  
%
%
Before commenting on this case in Sec \ref{Shape-Dyn}, I mention firstly that changing the assumed $\FrG$ from Diff($\bupSigma$) to just the trivial group consisting of the identity 
id gives a fifth prong, which can be encoded as {\sl unit-determinant} diffeomorphisms \cite{AM13}. 
Secondly, \cite{OM03} considered a furtherly structurally version of this working via considering a matrix rather than a metric, 
and then furthermore not presupposing the `spatial Ricci scalar combination of derivatives' of this matrix.

\subsection{Background independent aspects of the constructed spacetime}

If one chooses the GR prong of the fork, one recovers GR's notion of spacetime, and hence causal structure and the standard interpretation of multiple families of observers. 
Here then the conventional array of techniques used to study black holes are safe. 

\ni Step 12) The other independent extension is to follow construction of spacetime by envisaging this spacetime to possess its own form of Relationalism.
This is attained by the Lie derivative implementation of Diff($\FrM$). 

\ni Step 13) This constructed spacetime is then guaranteed to possess the Refoliation Invariance property.
This requires a Temporal Relationalism implementing foliation variant of Teitelboim's proof for spacetime assumed (see \cite{TRiFol} for details).

Note that the `spacetime from space' prong that the current Seminars concentrates upon additionally provides \cite{RWR, AM13} a second answer to Wheeler's question (\ref{Wheeler-Q}).
This goes beyond Hojman--Kucha\v{r}--Teitelboim's first answer's assumption of embeddability into spacetime.  
It achieves this by being based rather on 3-spaces from first principles in place of hypersurfaces which do presuppose a surrounding spacetime. 
Furthemore, it proceeds from Temporal and Configurational Relationalism first principles.
Then the consistency of the ensuing constraints' algebraic structure along the lines of the Dirac algorithm returns, from a more general $\scH_{\st\sr\si\sa\sll}$. 
GR's particular $\scH$ alongside local Lorentzian relativity and embeddability into GR spacetime here {\sl emerges} one of very few consistent possibilities. 
Thus Relationalism is not only a demonstration of the existence of a formulation in which GR is relational, but also, a fortiori, 
is its own route to GR (in Wheeler's sense \cite{Battelle}).

Also note that the present Sec complete understanding of the seven local aspects of Background Independence that lead to seven of the PoT facets at the classical and QM levels.

Finally, furthemore it transpired over the years that Temporal and Configurational Relationalism are highly unselective of standard theories, 
but the multi-aspect Background Independence is rather more selective.

\subsection{Shape dynamics alternatives}\label{Shape-Dyn}

This arises as the fourth prong of the fork (\ref{4-Factors}), 
or its subsequent encodement as a first-principles VPConf($\bupSigma$) or Conf($\bupSigma$) invariant theory \cite{ABFKO, Kos2, Mercati14, BO10}.
Here GR-like spacetime structure is not presupposed. 
This renders unclear the status of multiple families of observers (tied to refoliation invariance in the case of GR) and of causal structure. 
[Arguments assuming that a 3-metric, 3-vector and scalar can `just' be packaged together to make a 4-metric tend to fail in that they would also apply for such as Newtonian Gravity or 
Strong Gravity, which however are known {\sl not} to have such a `unified and simple' 4-geometry...]
This is part of why I for now choose the relational recovery of GR prong of the fork over the shape dynamics one. 
My position is in accord with looking to implement all types of Background Independence rather than just Temporal and Configurational Relationalism, and Constraint Closure 
(and perhaps then Expression in terms of Beables).
Furthermore, without these established to apply here, many elements used in the study of conventional GR-like black holes are open to question. 

\mbox{ } 

\ni Comment 1) Later, furtherly matured works on shape dynamics \cite{Kos2, Mercati14} use `symmetry trading' to pass between the symmetry corresponding to 
Refoliation Invariance and that corresponding to spatial (volume -preserving) conformal invariance. 
Symmetry trading is known from other areas of Theoretical Physics, 
and involves working in an extended phase space, within which the smaller initial and final theories both feature as particular sectors.
But if `symmetry trading' is to be applied to `shape dynamics', one had better check that the {\sl notion of symmetry} involved in such trading.
But the version they have developed so far rests upon the finite integral implementation, which firstly fails to implement $VPConf(\bupSigma)$ as a group, 
and secondly has generators lying well outside of the usual Lie theory. 
Thus it is far from clear whether such a generalized sense of symmetry lies within the scope of current treatises on `symmetry trading'.
Certainly the most commonly encountered examples of symmetry trading in Gauge Theory \cite{HTbook} do not by themselves suffice as a guarantee 
of being able to `trade' the far more general and far less explored notion of symmetry pointed to in this note.  
That is already the case even in the differential functional notion of constraint algebroid required to address GR.

\ni Comment 2) The $x = 1$ and $x \neq 1$ cases of shape dynamics likely have different statuses as regards Refoliation Invariance.
The first of these happens to fit the mathematical form by which GR attains Refoliation Invariance, whereas the second is not known to. 
But even for the first of these, if the CMC condition is afforded the status of a constraint rather than of just a particular slicing condition, 
what is the meaning then of a refoliation that takes one from a CMC foliation to a non-CMC one?

\ni Comment 3) Then also, in the particular case of shape dynamics, the incipient symmetry carries extra physical connotations that the final symmetry is not known to possess.  
This is because Refoliation Invariance is tied to how GR handles the multiplicity of possible families of observers. 
Thus the algebroid complication in the previous comment is tied intimately to the physical interpretation accorded to the theory.

\ni Question 1) Does trading Refoliation Invariance symmetry for spatial conformal invariance symmetry causes GR's manner of incorporating multiplicity of observers to be lost?
Or can a new and satisfactory way of encoding this involving spatial conformal invariance be found?

\ni Comment 4) Additionally, Refoliation Invariance also has the status of a hidden symmetry whilst spatial conformal invariance is manifest.
Thus one might also consider whether `symmetry trading' applies to trading {\sl hidden} symmetries for the usual manifest ones.

\ni Comment 5) A separate issue with symmetry trading concerns whether phase space extensions constitute mathematical super-structure. 
This is as opposed to adding no superstructures, and to attempting to remove conventionally used structures as putative superstructures in more actively minimalistic perspectives. 
One issue here is whether an extension introduced gives rise to a physical imprint.
In particular, does having CMC-sliced GR as but a sector within an extended phase space of shape dynamics allow for the {\sl other} sectors present therein causing a quantum level 
imprint on the physics?

\ni Question 2) If so, is this a source of insight and predictions, or is it just an uninvited import from the ad hoc adjunction of unphysical extra sectors?

\ni Question 3) Then what spacetime structure does shape dynamics possess, and subsequently or otherwise, 
what is the status of the accommodation of a multiplicity of families of observers and of causal structure in shape dynamics?  
\cite{GC} may be seen as a first step as regards addressing some parts of these questions.

\ni Furthermore, once these issues about spacetime structure, observers and causal structure are established, 
it can be determined whether some of the methods used to study black holes in theories with a GR-like spacetime notion remain applicable in shape dynamics.

\ni Comment 6) Whereas this comment originally concerned what Representation Thoery VPConf($\bupSigma$) would have, I first now need to point out that the implementation 
of volume-preserving conformal transformations usually used hitherto in the `shape dynamics' literature is not group-forming \cite{AConf2}.  
Various other more suitable implementations are discussed there, as is how \cite{BO10} manages to avoid working with CS + V altogether.
This is a more imminent problem for shape dynamics than the above comments and questions, 
since it points to a need to rework this from in a manner which has not lost contact with the intended underlying Group Theory.

Due to this, I consider discussion of Representation Theory of VPConf($\bupSigma$) to be premature. 
Of course, the point of Representation Theory is its use in QM.
A quantum treatment of CS would involve representations of 

\ni Conf($\bupSigma$) $\rtimes$  Diff($\bupSigma$). 
Since Conf($\bupSigma$) is well-known to be contractible, this is little different from Diff($\bupSigma$)'s own Representation Theory.
Koslowski has argued that shape dynamics is not amenable to simple expression in loop terms \cite{Kos13}.
Whereas one might wish to re-check this within the other schemes pointed to in the previous paragraph, 
if that is so, then one would be stuck with using the traditional representation theory of Diff($\bupSigma$), for which Isham and Kucha\v{r} pointed out many impasses \cite{I84IK85}.

\ni Comment 7) A number of shape dynamics papers ascribe significance to the York time \cite{YorkTime1} that bears close relation to the CMC condition.
Use of York time in earlier literature had a number of well-known shortcomings \cite{Kuchar81, I93, APoT2} though many of these do not necessarily apply in shape dynamics' new context. 
However, York time appears to be unconnected to what clocks actually read; indeed some shape dynamics paper involve two times: York time {\sl and} Machian emergent time. 
However, with clocks reading Machian emergent time, what it the York time needed for in this scheme?  
[On the one hand, York time appears to be {\sl useful for performing dynamical calculations}, 
whilst on the other hand there appears to be no {\sl operational significance} ascribed to this quantity.]

\section{Semiclassical extension of the Machian PoT strategy}

Within the `standard GR recovery' branch of the above, one can extend the Background Independence implementation and PoT tackling to quantum level.  
The {\it semiclassical approach} involves some heavy slow degrees of freedom provide an approximate emergent time with respect to which the other light fast degrees of freedom evolve.
See \cite{KieferBook, FileR, ACos2} for further details.  

\ni Facet 2) One hopes that classically-resolved Configurational Relationalism stays quantum-mechanically resolved (though anomalies are possible).  
This is the case for the RPM's considered, and is irrelevant in the case of minisuperspace
 
\ni Facet 1) The Wheeler--DeWitt equation's Frozen Formalism Problem still occurs and is not unfrozen by $t^{\se\sm(\sJ\sB\sB)}$. 
However $t^{\se\sm(\sW\sK\sB)}$ or $t^{\se\sm(\sr\se\scc)}$ can be abstracted from suitably semiclassical quantum change.

\ni Start afresh as regards obtaining an emergent time.
The classical--semiclassical distinction between these is itself well founded on Machian grounds: 
in the latter case {\sl quantum change} is to given the opportunity to contribute to the timestandard being abstracted from change.
See Fig \ref{Gates-3} for an outline of the procedure.
RPM's and minisuperspace considered have no quantum brackets issues, the slightly inhomogeneous cosmology case is slightly not checked yet.  

\ni Facet 3) Whereas we do not know how to handle quantum constraint closure in the case of general GR, for RPM's this is attained good fortune, 
whereas the minisuperspace case is aided by having only the one constraint.

\ni Facet 4) One either promotes one's classical level subalgebraic structure of \K beables to quantum operators or one start afresh at the quantum level.
 
\mbox{ }  
 
\ni Issue 1) Justifying the WKB regime is left open at this level; see Sec 5 for more.

\ni Issue 2) In the absense of being able to solve the classical (or semiclassical) Step 1), resolutions Steps 2) and 4) remain implicitly defined. 
We are only claiming a local resolution to apply to classical and semiclassical RPM and minisuperspace and, for now, classical-level slightly inhomogeneous cosmology.  

\ni Issue 3) For minisuperspace \cite{AMSS1} in comoving-type coordinates privileged by the surfaces of homogeneity, 
homogeneity provides a simpler resolution here of 6) and 7), both classically and quantum mechanically.
On the other hand, for RPM, facets 5)-7) are here unnecessary since these models do not possess a GR-like notion of spacetime.  
On the other hand, facets 3)-7) are nontrivially exhibited by perturbations about minisuperspace \cite{SIC1}, 
making that a good model for these aspects, especially at quantum level for which there is not a known resolution for the general GR case.

\ni If $\FrG$ is still quantum mechanically OK, there is no need for more Configurational Relationalism, and classical \K beables remain.

\ni On the other hand the statuses here of spacetime and of Refoliation Invariance are for now unestablished.  
One needs at least slightly inhomogeneous cosmology model arena to investigate these.  
The current Seminar is mostly a classical tour of the `gates of the castle'...  

{            \begin{figure}[ht]
\centering
\includegraphics[width=0.8\textwidth]{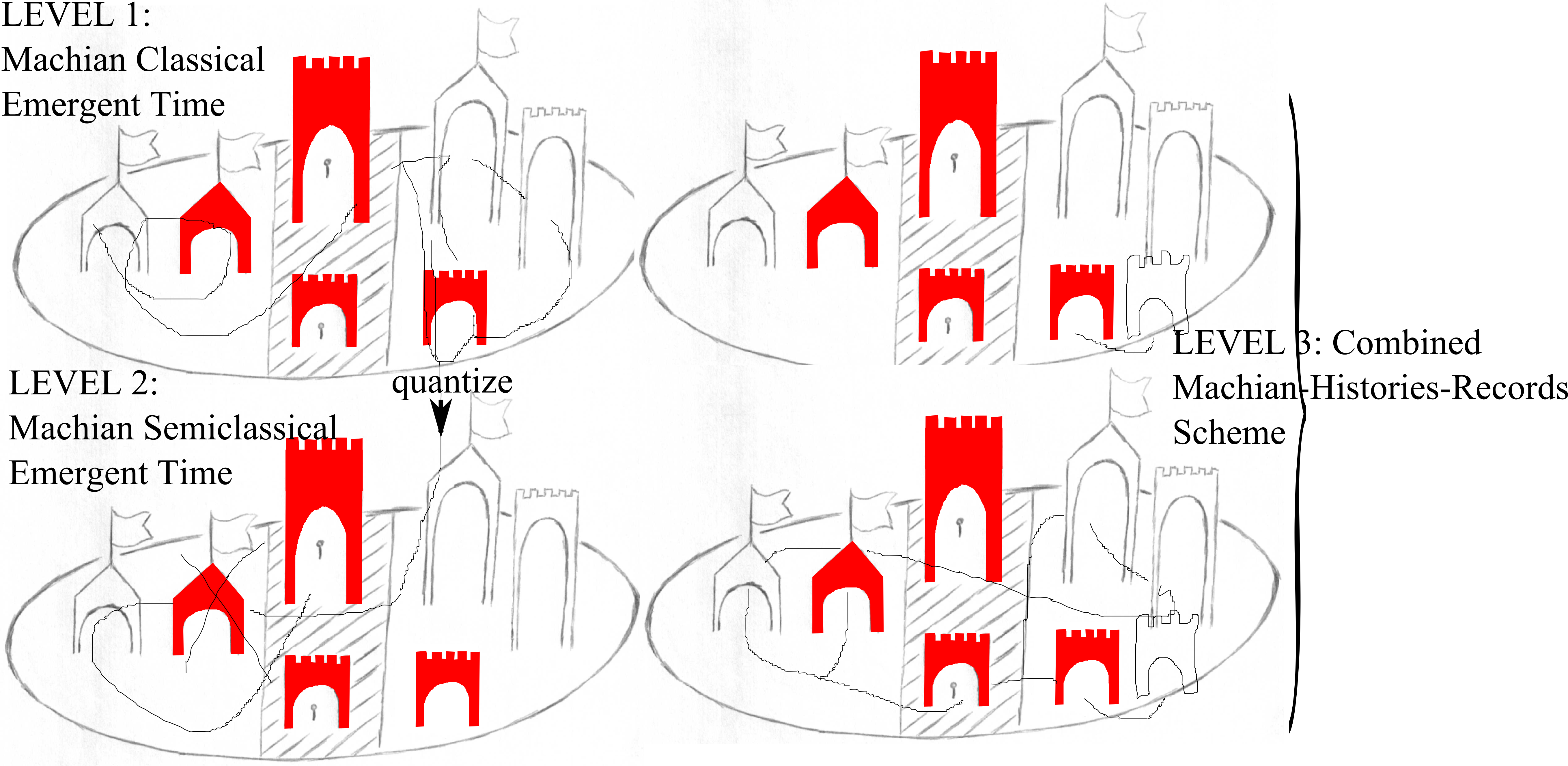}
\caption[Text der im Bilderverzeichnis auftaucht]{\footnotesize{a) Follows up on Fig \ref{Gates-2} with quantization and subsequent semiclassical consideration.
b) shows how to support this at the classical level with a further histories theory working.
The heavy line and its classical precursor are a further benefit of this combined scheme (I have now split the beables gate into constructions of \K and then Dirac beables).  
The Isham--Linden \cite{IL} formulation permits a classical precursor of the other histories moves, though these are much less required since decoherence is an entirely quantum notion.} } 
\label{Gates-3} \end{figure}          }
 
\section{Machian--Semiclassical--histories--records extension}

\ni Timeless Records involve considering only questions about the universe `being', rather than `becoming'. 
The {\it \CPI} for answering questions of conditioned being, which can then be about being at a time or about correlations.
Records \cite{PW83-B94II, GMH, H99} are localized subconfigurations of a single instant that contain information/correlations/patterns \cite{Records, ATop}.
In a purely timeless approach, these are useful as regards construction of a semblance of dynamics or history.

In {\it Histories Theory}, histories themselves are regarded as primary entities.  
Isham--Linden style histories \cite{IL} they possess their own conjugate momenta and brackets structure.
(The older Gell-Mann--Hartle \cite{GMH, Hartle} style histories are purely quantum).
N.B. that histories have a mixture of spacetime properties and canonical properties.

Then pairwise, one has I) Machian Records Theory.
II) Histories within the Machian time approach.  
III) The classical Records within Isham--Linden Histories Brackets \cite{IL} analogue of Gell-Mann--Hartle's better-known quantum inclusion of records within histories theory 
\cite{GMH, H99}. 
These two are additionally united as follows.

\ni Interconnection 1) Both histories and records fulfil Mackey's criterion \cite{IL, I10-ToposRev, FileR} by resting on atemporal logic.

\ni Finally, the triple combination is my Machianized $\FrG$-nontrivial \cite{AHall} of Halliwell's classical prequel \cite{H03}. 
The additional interprotection at the classical level is that the classical Machian approach or histories theory `provide a semblance of dynamics or history'.
This overcomes present-day pure records theory's principal weakness of not having well-established own means of providing such a semblance.

There is also a means of constructing classical Dirac beables in extension of Halliwell's \cite{H03, AHall, FileR} as a subset amongst the quantum \K beables. 
This involves classical timeless probabilities for histories entering a region $R$ of configuration space.

Concurrent deployment of some combinations of the individual strategies shows capacity to remove weaknesses from each.
Most of the value of the combined approach, however, is at the semiclassical quantum level.  
Here the additional interprotections are as follows. 

\ni Interprotection 2) The basic idea is to prop up the principal deficiency of Level 2 -- justification of the assumption of a a WKB regime -- 
using decoherence in the form of histories decohereing \cite{ZehBook, KieferBook}. 

\ni Interprotection 3) As Gell-Mann and Hartle said \cite{GMH} 

\beq
\mbox{``{\it records are somewhere in the universe where information is stored when histories decohere}" }.
\eeq

\ni Interprotection 4) One can answer the elusive question of `what decoheres what' from what the records are.

\ni Interprotection 5) By providing an underlying dynamics or history, whichever of the semiclassical Machian scheme or the histories scheme 
overcome present-day purely timeless records theory's principal weakness of needing to find a practicable construction of a semblance of dynamics or history.

\ni Interprotection 6) The Semiclassical Approach provides a Machian scheme for quantum histories and quantum records to reside within. 

\ni Interprotection 7) The semiclassical regime aids in the computation of timeless probabilities (see below for more).

\ni At the classical level, Interprotections 2--4) are absent since they concern the purely quantum notion of decoherence, 
and Interprotection 7) vanishes since it concerns a purely quantum probability computation. 
At the quantum level, Interprotection 1) is far more significant than at the classical level too: standard logic versus Topos Theory's nontrivial intuitionistic logic \cite{I10-ToposRev}.

How does the combined scheme fit together as regards primality?  
{\sl Meaningless label} histories come first; these provide the regime in which the Semiclassical Approach applies and then this in turn gives the version of the histories approach 
in which the histories run with respect to Machian semiclassical emergent time.
Then localized timeless approaches sit inside the last two of these schemes. 
On the other hand, the Semiclassical Approach sits inside the global timeless approach. 
However the global timeless approach can be taken to sit within global meaningless label time histories approach.
Thus down both strands of the argument, histories are the most primary entities in the combined approach.

Returning to Interprotection 7), this additionally provides `start afresh' means of construction of semiclassical Dirac beables in extension of Halliwell's 
\cite{H03, AHall, FileR} as a subset amongst the quantum \K beables.
This involves a type of histories-theoretic {\it class functional}.

Thus preliminarily start again with each of histories and records at the classical level, since we will be combining these with Machian classical and semiclassical approaches.  
Consider histories with respect to label time as a construct to be jettisoned later.
From decoherence here, obtain emergent time. 
Then consider Machian emergent time histories, making the Machian version of Halliwell's procedure \cite{AHall, A13, FileR, ABook}.  
This mutually supporting semiclassical--histories--records perspective, minus its Machian element, was developed by Halliwell \cite{H03}. 
Finally, I then showed \cite{AHall, A13} that its Machian counterpart is a natural third-generation Machian strategy 
following on from classical and then semiclassical emergent Machian time.

\section{Conformal, supersymmetric and topological extensions}

One can invest in questioning whether scale is an undesirable absolute structure.
On the GR side of Fig 1's cube, this could be done at the level of spacetime conformal scalings or of space ones, 
whilst on the QFT side Conformal Field Theory (CFT) is a natural and well-documented option.

However, this is not unique as a natural first choice of alternatives to the Planckian cube of theories; another such is Supersymmetry (Sec \ref{Sugra}). 
This is also natural and well-documented on the QFT side, and arises within a pure GR approach also from asking about the form of $\scH$ \cite{SqrtTeitelboim}
(Supergravity as the `square root' of $\scH$, much as Dirac Theory is of Klein--Gordon Theory).
See the next Sec for on outline of this.

Finally, one can question whether the fixed spatial topology that pervades Geometrodynamics and Loop Quantum Gravity is an undesirable absolute structure.  
On the QFT side of the cube, Topological Field Theory (TFT) -- particularly metric-free versions such as Chern--Simons Theory \cite{TFT} -- 
is natural as a second variant following on from CFT.  
On the GR side of the cube, consideration of topology change in GR \cite{Top-Change} began with Wheeler's envisaging of spacetime foam \cite{WheelerGRT, Wheeler64b, Battelle}.  
Considering not only the metric level but also topological manifold level Background Independence originates in these works.
In incorporating topology change into geometrodynamics, singular metrics need to be included and one considers the Lorentzian-signature version of cobordisms between spatial topologies.  
See Sec \ref{Deeper} for more on Background Independence for topological manifolds and at yet deeper levels of structure.

\section{Supergravity's Background Independence is different}\label{Sugra}

Assume the spacetime position for the Supergravity action, apply the space-time split, and cast that in canonical terms (i.e. the Supergravity parallel of the path in Fig 6.c).
In this way, one obtains constraints $\scH$, $\scM_i$, $\scJ$ and $\scS$ 
(suppressing spinorial indices and complex conjugations for simplicity of presentation, since these are not needed for the below overview; see e.g. \cite{DEath-VM} for explicit forms).
The $\scJ$ occur in first-order formulations, as necessitated by the inclusion of fermionic variables.
Supersymmetric theories then have additional supersymmetric constraints $\scS$.

Next, upon considering Constraint Closure (see \cite{DEath-VM} for details), a qualitatively distinct feature from GR becomes apparent: schematically,

\ni \beq
\mbox{\bf \{}\scS \mbox{\bf ,} \, \scS \mbox{\bf \}} \mbox{ } \mbox{ } \widetilde{\mbox{ }} \mbox{ } \mbox{ } \scH + ... 
\label{SS-H}
\eeq
\mbox{ } \mbox{ } Moreover, (\ref{SS-H}) has consequences for how `constraint providers' can be regarded \cite{ABeables, AMech}.
Firstly, it is a statement that linear constraints do not in this case form a subalgebraic structure.  
This signifies a breakdown in Best Matching being the provider of linear constraints.

On the other hand, {\it linear bosonic constraints} $\scB\scL\scI\scN$ := $\scM_i$, $\scJ$ {\sl do} form a subalgebra 
(though making such a block-identification runs against the grain of supersymmetry).   
Thus Best Matching can provide these, leaving one needing constraint providers for $\scH$ and/or $\scS$.  
This might involve a new or modified constraint provider. 
(\ref{SS-H}) might be taken to signify that $\scH$ provision -- by Temporal Relationalism -- 
is to be supplanted by $\scS$ provision based on supersymmetric first principles, since providing $\scS$ suffices to obtain $\scH$ also, by the Dirac algorithm.

One can additionally think of (\ref{SS-H}) as $\scS$ being a construction of a square root of $\scH$  \cite{SqrtTeitelboim}
                               in parallel to how the Dirac equation arises as a square root of the Klein--Gordon equation.  
Thus Supergravity may provide reasons why $\scQ\scU\scA\scD$ or $\scC\scH\scR\scO\scN\scO\scS$ constraints are, after all, not so fundamental. 
Supergravity may merit {\sl shifting} Wheeler's question (\ref{Wheeler-Q}) from concerning $\scH$ to concerning $\scS$: 
why this takes the form it does (and thus what is its underlying `zeroth principles' or constraint provider. 
It should however be cautioned that whereas the fermions corresponding to Dirac's square root were subsequently observationally vindicated, 
this is not the case to date as regards superpartner particles.  
This can be taken as a limitation on arguing against the fundamentality of quadratic constraints like $\scH$ on the grounds of their being supplanted in supersymmetric theories.

Perhaps then Supergravity is just based on spacetime-primary formulations.  
Best Matching is not itself a supersymmetric concept, for all that it can still be applied to the bosonic sector of supersymmetric theories.

A second consequence of (\ref{SS-H}) is that there is no supersuperspace (Supergravity analogue of Wheeler's superspace).
There is a non-supersymmetric superspace obtained by quotienting $\scB\scL\scI\scN$ out.

Compared to theories preceding it, Supersymmetry is conceptually strange in how the product of two supersymmetry operations is well-known to produce a {\sl spatial} transformation 
In other words, this theory mixes spatial and internal transformations.
I.e. {\sl The standard picture of Relationalism lies within the auspices of the Coleman--Mandula Theorem, whereas supersymmetry is the prime means of eluding this No-Go Theorem.}
This is in the sense of there being a well-defined notion of Configurational Relationalism which factors into spatial and internal parts; 
Spacetime Relationalism factors into spacetime and internal parts too.

For GR, the canonical formulation can be factored into separate space and time pieces via $\scM_i$ and $\scH$ admitting a pure-stretch of space itself and a 
pure deformation of space amounting to its time evolution, with pure stretches closing as an algebra involving space alone. 
These constraint actions still hold in Supergravity, 
but the presence of the additional $\scS_A$ breaks the `mandate' of $\scF\scL\scI\scN$ acting within configurations and $\scQ\scU\scA\scD$ altering configurations.

Various possible alternatives are as follows.

\ni a) Supersymmetry is one of a) not conceptually sound, {\sl or} 

\ni b) Supersymmetry is conceptually deeper than many other transformations, requiring development of the theory of Relationalism and Background Independence beyond that provided 
in this Seminar. 

\ni I.e. {\sl the Relational Program might eventually provide its own prediction as to whether to expect Supersymmetry in nature. 
Or, conversely, insisting on Supersymmetry could provide a new direction in which to generalize the current conceptualization of Relationalism and Background Independence.}    
See \cite{AMech} for an update.

I have also set up the supersymmetric analogue of RPMs \cite{AMech}. 
I have then noted that ease of reduction of examples of \cite{BB82} transcend to the bosonic sector of the supersymmetric theory, and yet leave the fermionic sector in absolute form.

\ni As a third consequence, 
$\scF\scL\scI\scN$ not forming a subalgebraic structure implies \K beables are not well-defined for Supergravity by the Casalbuoni brackets extension of working (\ref{CCB}).  
Non-supersymmetric -Kucha\v{r}: beables are well defined, but are not a supersymmetric notion.

Finally, the possibility of Spacetime Reconstruction in Supergravity is beyond what this Seminar can consider.

All in all, by this stage Supergravity is revealed to be far more classically-distinct from Geometrodynamics than Loop Quantum Gravity is.
The current Relationalism and Background Independence program can readily be extended to the latter but not to the former!
Consequently, the PoT is substantially different for Supergravity, 
rendering it an arena of considerable interest for future investigations into the nature of time and the foundations of Quantum Gravity.

\section{Background Independence and the PoT for the deeper levels}\label{Deeper}

Isham's \cite{I89-Latt, IKR-I91, I03, I10-ToposRev} pioneered allowing a number of the deeper levels of structure that are outlined in Appendix A to be quantized.
I am laying out dynamics and stochastic schemes as classical preliminaries for these in \cite{ATop, AMech, ABook} for both GR and for RPM model arenas.
Here Relationalism has received more attention than Background Independence, which has received more attention in the Physics literature than associating stochastic mathematics.

Some relational considerations for this are as follows.
Temporal Relationalism descends the levels of structure: each can have no time for the universe as a whole, whilst possessing a notion of change from which time can be abstracted.
Configurational Relationalism also descends the levels of structure: each can be associated with a group and taken into account with an $\FrG$-act--$\FrG$-all pair.  
The Appendix lists natural choices of groups to be physically irrelevant at each level of structure.

Next note that generalized phase spaces can be ascribed, whence generalized momenta and generalized brackets.  
Some classical level limitations from breakdown of differentiability, but the quantum commutators remain defined.
In this setting notion of constraint retains meaning all the way down, as does issue of entities forming zero brackets with the constraints.

There are distinct spacetime and space floors for each of the levels of structure. 
There is a progressive loss of distinctions between spacetime and space as one descends the levels of structure: signature is lost, 
then codimension 1 becomes meaningless with loss of dimension, leaving space being a strict subset of spacetime.
I use `space', `time', `spacetime', `slice', `foliate', `surround', `construct' ... as level-independent concepts.
{\sl This conceptualization indeed points to many further versions of the PoT facets at each of these levels of structure.}
  
\mbox{ } 

\ni As regards timeless records, Kendall \cite{Kendall} has given an account of probing 2-$d$ $N$-point constellations probed by triples, 
in terms of geometrical probability theory on the shape sphere \cite{Kendall}.  
In fact, he further considered the more complicated and general case of geometrical probability on stratified manifolds \cite{Kendall}, as is required in the 3-$d$ case.  

At the deeper level of the lattice formed by the topologies on a fixed set, techniques developed by Molchanov \cite{Molchanov} may apply. 
This is furthermore a setting in that Isham long previously considered the quantization of \cite{I89-Latt, IKR-I91}.  
In some of these works, he also considered the simpler case of quantizing spaces of metric spaces.  
Even earlier, Kendall developed a stochastic theory of collections of random sets \cite{Kendall73}. 
See \cite{ATop} for yet further examples.

\mbox{ } 

\ni Isham also pointed out a limitation in the endeavour of considering spaces of spaces for Theoretical Physics, namely that Russell's paradox applies to the set of sets

\mbox{ } 

\ni Finally, equipped sets is not unique as a structural paradigm for mathematics as a whole.
Whilst for now \cite{ATop} concentrates on this, other alternatives include Category Theory, Topos Theory and the Synthetic approach, 
with Isham so far having started consideration of quantization of the first two \cite{I03, I10-ToposRev}.  
Indeed, finally, do Relationalism or Background Independence more generally shed any light on {\sl this} choice? 
A simple observation here is that the category of categories also suffers the above limitation, 
but e.g. the category of small categories (as used in \cite{I03}) does not since it is not itself small.  

\mbox{ } 

\ni {\bf Acknowledgements}: 
E.A. thanks close people. 
The Conference Organizers for the invitation to speak here and hospitality. 
The Conference participants and Chris Isham for discussions.
Jeremy Butterfield, John Barrow, Marc Lachi$\grave{\me}$ze--Rey, Malcolm MacCallum, Don Page, Reza Tavakol and Paulo Vargas-Moniz for help with my career.

\begin{appendix} 

\section{Levels of mathematical stucture very briefly explained}

\ni Begin by perusing Fig \ref{Bigger-Set-2}.

{            \begin{figure}[ht]
\centering
\includegraphics[width=0.8\textwidth]{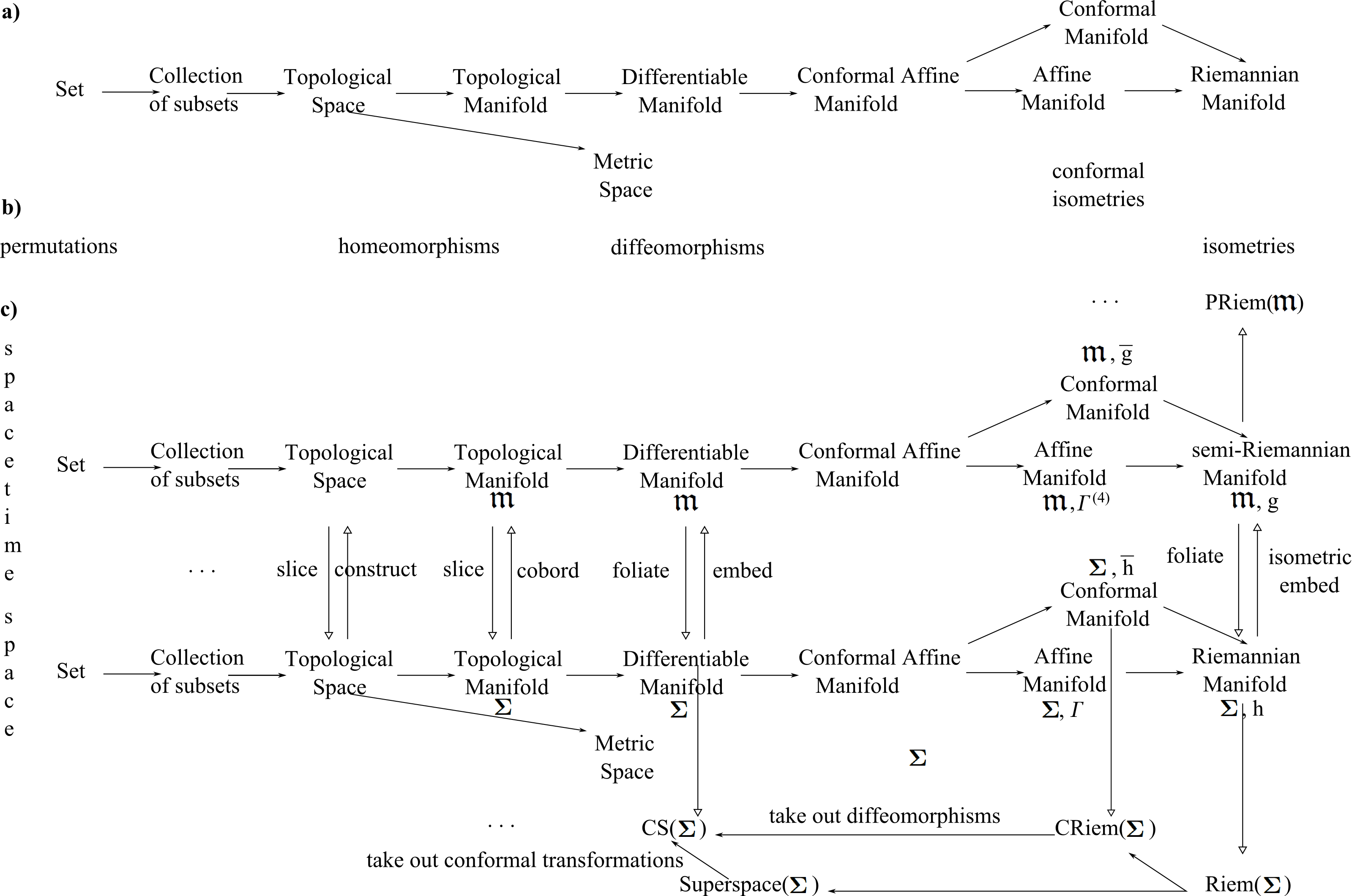}
\caption[Text der im Bilderverzeichnis auftaucht]{        \footnotesize{a) Levels of mathematical structure commonly assumed in Classical Physics.  
\cite{ATop} considers allowing for each of these to be dynamical in turn.  
Note that `metric structure = conformal structure + localized scale', and then in the indefinite spacetime metric case, conformal structure additionally amounts to causal structure.   
b) Each level's particularly significant morphisms.
c) Specific spacetime and space versions, interlinked with maps and including corresponding spaces of spaces 
(I only include some of them as the article does not have the space to introduce the notation for the rest of them).
Note in particular the advent of stratified manifolds.} }
\label{Bigger-Set-2}\end{figure}            }

\ni A {\it topological space} is a set $X$ alongside a collection of open subsets $\tau$ such that 

\ni T1) The union of any collection of these subsets is also in $\tau$.

\ni T2) The intersection of any finite number of these subsets is also in $\tau$.  

\ni T3) $X, \emptyset \in \tau$.

\ni However, it is not a priori clear that one wishes to consider this kind of collection, 
as is clear from e.g. $\sigma$-algebras, filters \cite{AMP} and trapping sets \cite{Kendall73} being other kinds of collection. 
It is due to this that I include `collections of subsets' as an intermediate step in Fig \ref{Bigger-Set-2}.a). 
Do Relationalism or Background Independence shed any light on {\sl this} choice? 

\mbox{ } 

\ni Next, a   {\it topological manifold} is then a topological space that is 

\ni M1) {\it Hausdorff}:        for $x, y \in X, x \neq y$, $\exists$ open sets $O_x, O_y \in \tau$  such that $x \in O_x, y \in O_y$ and $O_x \bigcap O_y = \emptyset$. 

\ni M2) {\it Second-countable}: there is a countable collection of open sets such that every open set can be expressed as union of sets in this collection. 

\ni Note that these axioms perform a selection of the middle ground, 
much like Goldilocks rejecting the porridge that is too sweet as well as that which is not sweet enough, before finding a bowl that is `just right'.  
I.e. Hausdorffness guarantees not too small and second-countability guarantees not too large. 
Though these senses of small and large are not exact science, much like porridge-making is not. 
Thus one might consider a somewhat wider range of topological manifolds with middling properties by considering slightly different countability and separation axioms 
(Hausdorffness being an example of the latter).
Do relationalism or background independence shed any light on {\sl this} choice? 

\ni M3) {\it Locally Euclidean}: every point $x \in X$ has a neighbourhood $N_x$ that is homeomorphic to the Euclidean space $\mathbb{R}^p$ (meaning we can use charts).  

\ni Topological manifolds are an extension of the notion of continuity; extending differentiability in a similar vein specializes these to differentiable manifolds. 
Here the chart concept is used to everywhere locally harness $\mathbb{R}^n \rightarrow \mathbb{R}^n$ function calculus.  
One can further equip these with affine structure, conformal structure and metric structure in the sequence indicated in Fig \ref{Bigger-Set-2}.a).

\mbox{ } 

\ni {\it Stratified manifolds} in general lose the three manifoldness axioms, but many relevant examples fortunately retain the familiar Hausdorfness and second countability properties. 
They comprise a set of manifolds (not necessarily of the same dimension, so local Euclideanness is certainly lost), 
which are none the less fitted together in a relatively benign manner by obeying the {\it frontier property}. 
Namely that, for any two of the manifolds, say $\FrMgen$, $\FrMgen^{\prime}$ with $\FrMgen \neq \FrMgen^{\prime}$

\ni
\beq
\mbox{ if } \FrMgen^{\prime} \bigcap \overline{\FrMgen} \neq \emptyset \mbox{ } , \mbox{ } 
\mbox{ then } \FrMgen^{\prime} \subset \overline{\FrMgen} \mbox{ and } \mbox{ } \mbox{dim}(\FrMgen^{\prime}) < \mbox{dim}(\FrMgen) \mbox{ } .
\eeq
We then say a partition into manifolds has the {\it frontier property} if the set of manifolds has.
Finally, a {\it stratification} \cite{Whitney65} is a strict partition of one's topological space which has the frontier property.
%

At the level of spaces of spaces, note that manifold and stratified manifold concepts translate to \cite{Fischer70} cases in which $\mathbb{R}^n$ 
is replaced by suitable infinite-dimensional linear spaces such as Banach spaces or Fr\'{e}chet spaces \cite{AMP}.  

\mbox{ } 

\ni Finallyt, the topologies on a fixed set form a lattice.
Introduce first a {\it poset} (partially ordered set): a set $X$ equipped with a {\it partial order}: 
a reflexive, antisymmetric and transitive binary relation (e.g. $\subset$, or $\geq$). 
Then a {\it lattice} is then a poset within which each pair of elements has a least upper bound and a greatest lower bound;   
in the context of a lattice, these are called {\it join} $\lor$ and {\it meet} $\land$.  
For the lattice of topologies on a fixed set, the partial order in question is the relative coarseness of the topologies in question.  

\end{appendix} 


\end{document}